\documentclass[journal]{IEEEtran}

\usepackage{amsmath}
\usepackage{multicol}
\usepackage{amsfonts,amssymb} 
 \usepackage{mathrsfs}
 \usepackage{amsthm}
 \usepackage{subfigure}
 \usepackage{graphicx}
    \usepackage{tabularx}
     \usepackage{supertabular}
     \usepackage{booktabs}
      \usepackage{threeparttable}
      \usepackage{multirow}

\usepackage{stfloats}
\usepackage{comment}
\usepackage{cite}
\usepackage{ifpdf}
 \ifpdf
 \else
 \fi

\usepackage{algorithmic}
\usepackage{array}

  \usepackage[caption=false,font=normalsize,labelfont=sf,textfont=sf]{subfig}

\ifCLASSOPTIONcaptionsoff
  \usepackage[nomarkers]{endfloat}
\let\MYoriglatexcaption\caption
 \renewcommand{\caption}[2][\relax]{\MYoriglatexcaption[#2]{#2}}
\fi
\usepackage{url}

\begin{document}
%
\title{Privacy Inference Attacks and Defenses in Cloud-based Deep Neural Network: A Survey}
%
%
%

\author{Xiaoyu Zhang,~\IEEEmembership{Member,~IEEE,}
        Chao Chen,~\IEEEmembership{Member,~IEEE,}
         Yi Xie,~
        Xiaofeng Chen,~\IEEEmembership{Senior Member,~IEEE,}
        Jun Zhang,~\IEEEmembership{Senior Member,~IEEE,}
        and~Yang Xiang,~\IEEEmembership{Fellow,~IEEE}

\thanks{Xiaoyu Zhang is with the State Key Laboratory of Integrated Service Networks (ISN), Xidian University,
	Xi'an, 710071, Shaanxi and 
	the State Key Laboratory of Cryptology, PO Box 5159, Beijing 100878, China (e-mail: xiaoyuzhang@xidian.edu.cn)}

\thanks{Chao Chen is with the College of Science and Engineering, James Cook University, Douglas, QLD, Australia (e-mail: chao.chen@jcu.edu.au)}

\thanks{Yi Xie is with the State Key Laboratory of Integrated Service Networks (ISN), Xidian University,
	Xi'an, 710071,  China (e-mail: yixie1997@gmail.com) (Corresponding author)} 

\thanks{Xiaofeng Chen is with the State Key Laboratory of Integrated Service Networks (ISN), Xidian University, Xi'an, 710071, Shaanxi,  China (e-mail: xfchen@xidian.edu.cn)}

\thanks{Jun Zhang is with the School of Software and Electrical Engineering, Swinburne University of Technology, Hawthorn, VIC 3122, Australia  (e-mail: junzhang@swin.edu.au)}

\thanks{Yang Xiang is with the School of Software and Electrical Engineering, Swinburne University of Technology, Hawthorn, VIC 3122, Australia  (e-mail: yxiang@swin.edu.au)}
}

%
%

\markboth{Journal of \LaTeX\ Class Files,~Vol.~14, No.~8, August~2020}%
{Shell \MakeLowercase{\textit{et al.}}: Privacy Inference Attacks and Defenses in Cloud-based Deep Neural Network: A Survey}
%



\maketitle 
\begin{abstract}
Deep Neural Network (DNN), one of the most powerful machine learning algorithms, is increasingly leveraged to overcome the bottleneck of effectively exploring and analyzing massive data to boost advanced scientific development.
It is not a surprise that cloud computing providers offer the cloud-based DNN as an out-of-the-box service. Though there are some benefits from the cloud-based DNN, the interaction mechanism among two or multiple entities in the cloud inevitably induces new privacy risks. This survey presents the most recent findings of privacy attacks and defenses appeared in cloud-based neural network services. We systematically and thoroughly review privacy attacks and defenses in the pipeline of cloud-based DNN service, \textit{i.e.}, data manipulation, training, and prediction. In particular, a new theory, called cloud-based ML privacy game, is extracted from the recently published literature to provide a deep understanding of state-of-the-art research. Finally, the challenges and future work are presented to help researchers to continue to push forward the competitions between privacy attackers and defenders.
\end{abstract}


\begin{IEEEkeywords}
Privacy Inference Attack, Privacy Defense, Deep Neural Network, Cloud Computing
\end{IEEEkeywords}

%
\IEEEpeerreviewmaketitle

\section{Introduction}
%
%
%
%
\IEEEPARstart{W}{ith} the increasing demand of artificial intelligence and advancement of computing hardware in the past decade, machine learning has experienced the most rapid growth in both academia and industry. Among the machine learning algorithm families, Deep Neural Networks (DNNs) are the most powerful and popular set of machine learning algorithms. A DNN is a generalization of regression to learn sophisticated relationships between high-dimensional input feature vectors and predictions.
It can offer intelligent decision-making and reasoning if fueled with a large amount of data \cite{chen2, yuantian2021, coulter2,  lin,  liu2, liu3, liu4, sun, zhang2, zhang3}. More recently, DNNs have been extensively used in various domains, e.g., image processing \cite{he, krizhevsky1, simonyan1}, natural language processing \cite{Goldberg}, speech recognition \cite{Amodei, Graves,  Hannun, Hinton, Toshev}, \textit{etc.}, and resulted in breakthroughs in these areas. However, DNN is computationally intensive, whose training process requires high computational resources. Due to the limited computing resources offered by the off-the-shelves workstations, DNN is infeasible to be applied at the end user's side. Moreover, the lack of massive public datasets will lead to low performance and overfitting of the machine learning model. Fortunately, the cloud-based neural network has become a hot and cutting-edge technological offering and attracted much attention from both the private and public sectors. Cloud computing \cite{ chen5, chen6, chen7, wang1, wang2, zhang4} makes it possible to allow the Internet users to employ the computing resources and scalable storage from a cloud service vendor and realizes the need of training a large neural network. In the cloud-based neural network paradigm, users with limited computing power are capable of outsourcing the heavy training tasks to the service vendor and leverage the rich resources in a pay-as-you-go manner. 

Despite the benefits, the paradigm of cloud-based neural network training \cite{ma1, ma2, zhang5, zhang6} inevitably introduces some new privacy issues \cite{minghao2020}. The first one is the training data privacy. Since the training datasets from data owners are often sensitive, for instance, purchase history records, engineering data and individual health records, it is necessary to guarantee the confidentiality of training data. The second one is the trained models' privacy. Neural network models are trained by learning from sensitive datasets, and consuming considerable computational resources or deployed for security applications so that it can be deemed as a type of intellectual property due to its commercial value. The third one is the query data privacy. When the model consumers send the request to the service providers to classify new samples, it poses the querying data and classification results at privacy risks. 
There are many research works done in the three common categories of privacy issues during neural networks learning and applications. 

Without loss of generality, the workflow of a neural network consists of three phases according to the sequence of events, namely \emph{Data manipulation}, \emph{Training}, and \emph{Prediction}. In this survey, we systematically and thoroughly explore the privacy issues occurring in these three phases from the attacker's and defender's view, respectively, \textit{i.e.,} privacy inference attack and privacy protection. Specifically, there are four types of malicious attacks on privacy, 1) membership inference attack: given a data record, one attacker aims to determine whether it comes from the training dataset; 2) property inference attack: it is launched by an attacker to infer some fundamental statistical properties, preserved by the participants’ training datasets; 3) model inversion attack: it refers to deduce training classes against neural network via reproducing a representative data record for the target category (training class inference) or recovering the training data record (data reconstruction); 4) model extraction attack: an attacker with black-box or grey-box access with no prior knowledge of the structures of the models or parameters targets at copying the functionality of this model, \textit{i.e.,} stealing the target model. On the contrary, a large body of works focus on protecting privacy concerning all phases of neural networks, leveraging differential privacy, adversarial machine learning, watermarking techniques and cryptographic techniques. Their main goals are identical to each other, protecting data privacy or alleviating information leakage from the model or its outputs.

There are some related surveys on security and privacy of machine learning.
Papernot et al. \cite{Papernot} reviewed ML security and privacy attacks within the adversarial machine learning framework. Some mitigation strategies were presented from a perspective of passive defense. Gong et al. \cite{gong1} presented a relatively comprehensive survey on incorporating differential privacy with machine learning.
Maria et al. \cite{Maria} focused on surveying attacks against privacy and confidentiality of machine learning algorithms. 
Recently, a large number of researchers have devoted themselves to researching ML privacy attacks and defense strategies orthogonally. 
The application of cloud-based neural network has been demonstrating its excellent market value and attracting many translation research.
We observe there is a lack of a dedicated survey for the privacy over the cloud-based neural network. 
This motivates us to collect, analyze, and review recent high-quality papers in the new topic. 
Our survey reviews existing attack and defense techniques across the workflow of cloud-based neural network.

We searched the research papers from high-quality cyber security conferences and journals ranging from 2015 to 2020, \textit{e.g.},  ACM CCS, NDSS, IEEE S\&P and USENIX Security. The main keywords used to seek these related papers are ``neural network'',  ``privacy attack'', and ``privacy defense''. Then, we filtered out some research papers irrelevant to the cloud. It is to restrict the surveyed scenarios to the deployed DNN models in a cloud environment. We further refined the paper collection by manually checking the quality of all papers. Finally, 32 high-quality papers on privacy attack/defense in cloud-based neural network applications were selected and reviewed in this survey.
  
This paper surveys privacy attack and privacy defense involved in the cloud-based neural network from a new perspective and catches up with the trend of privacy issues in a deep neural network. Our contributions are summarized as follows:
\begin{itemize}
    \item We introduce the privacy attacks targeting cloud-based Neural Networks, which aim at the confidentiality of training data and model internals (\textit{e.g.}, parameters or hyperparameters), along with their defenses. We provide a taxonomy of privacy attacks and defenses based on the phases of the occurrence in the workflow of the cloud-based neural network, \textit{i.e.}, \textit{Data manipulation, Training and Prediction}.   
    \item We systematically and thoroughly review the papers on privacy attacks and defenses in the past five years and analyze the game between neural network privacy attackers and defenders. We also classify the attacks into four types based on the targeted information, \textit{i.e.}, \textit{property inference attack,  model extraction attack, model inversion attack},  \textit{membership inference attack}. Besides, we group the defense approaches into four categories, including \textit{differential privacy, adversarial machine learning, watermarking technique} and \textit{cryptographic technique}.
    \item We discuss the challenges of privacy attacks on cloud-based neural networks and forecast the future attack models. From the defenders' view, we also discuss the challenges of how to defend against emerging privacy attacks and point out some potential future directions.
\end{itemize}
The remaining of this survey is organized as follows: the research methodology concerning to privacy attacks and their defenses is introduced in Section II. In Section III and Section IV, we review the literature of privacy attacks, and their defenses occurred in the three phases of cloud-based DNN pipeline in the past five years. In Section V, we discuss the challenges and point out some future work. Finally, a conclusion of this survey is presented in Section VI.


\section{Taxonomy and Methodology of Privacy in Neural Network}

Privacy has been defined as a human right and is referred to as ``the right to be let alone” \cite{warren} in 1890. This generic definition is not suitable for today's digitally-connected world where a large volume of personal data are generated, aggregated, stored, processed and utilized. Thus, it makes defining privacy difficult. The General Data Protection Regulation (EU) 2016/679 (GDPR) is a European regulation on data protection and privacy for all individual citizens of the European Union (EU) and the European Economic Area (EEA). The GDPR was accepted on 14/04 2016 and became enforceable from 25/05 2018. GDPR provides seven privacy principles including purpose specification, data collection and minimization, accuracy, limitation of use, accountability, information security, openness and transparency. \cite{Voigt}
In this survey, we mainly concentrate on privacy concerning the cloud-based DNNs pipeline, referring to the model and its data. More concretely, we take the life cycle of the ML system from training to prediction phase into consideration and analyze the adversary's main objectives and strategies at every phase. We observe that attacks aim to expose the private information about training datasets (data source) collected from the Internet user, query data submitted by model users and the model structure, including parameters, and hyperparameters (which can be regarded as intellectual property). On the contrary, the protective or defensive measures target achieving the requirement that the information leaked by an ML working chain should be limited to that from its calculation outputs.

\begin{figure*}[!thbp]
	\centering
	\includegraphics[width=.5\textwidth]{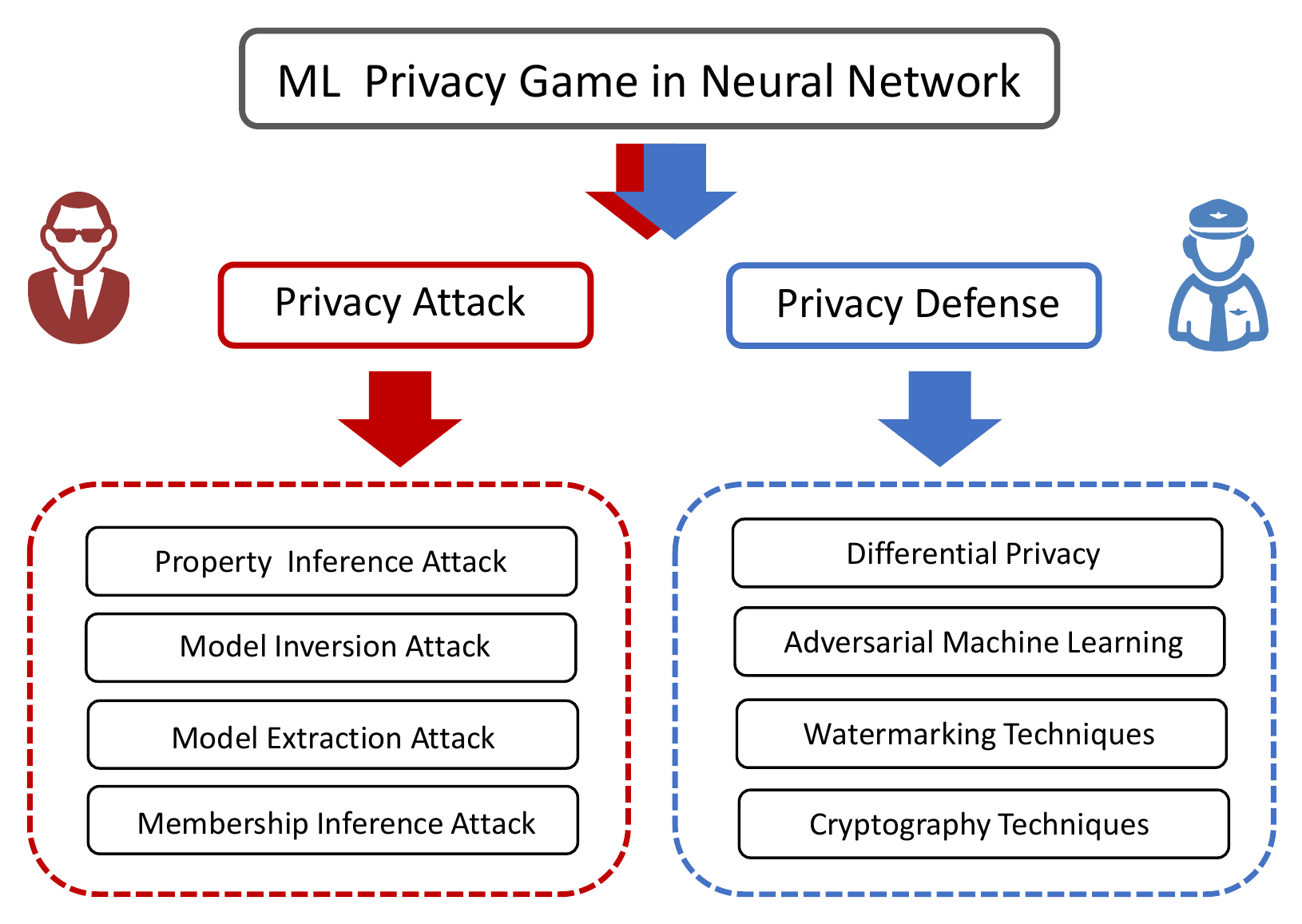}
	\caption{Taxonomy of ML privacy game in neural network. We classify these privacy issues into two categories in terms of privacy attack and defense. Depending on the type of information leakage, it includes property inference attack, model extraction attack, model inversion attack and membership inference attack. Meanwhile, data privacy protection strategies can be divided into four categories when using different techniques, \textit{i.e.,}  differential privacy,  adversarial machine learning, watermarking techniques and cryptography techniques.}
	\label{figure_system_model}
\end{figure*}



In real-world applications, one client may want to obtain a highly-accurate model with the assistance of a powerful cloud server. Several clients may work together to jointly train a more accurate deep neural network model. In general, the data owned by an individual often contains some sensitive personal information, such as their habits, photos and locations. To solve the privacy concerns during model training, it allows no potentially sensitive information to be disclosed to others, including the cloud service provider and other participants.
Thus, tremendous efforts are required for the design of secure protocol to collect these distributed data to constitute a centralized training dataset. This process is named ``data manipulation''. So far, the foreword is ready, we move on to training phase with the attempt to completely explore these aggregated datasets. In most cases, if there only exists one server training on a set of encrypted data guaranteed by CCA (Chosen-Ciphertext Attack) security, it is not necessary to discuss privacy issues in this phase. Hence, we concentrate on collaborative ML training scenario in the training phase. This scenario indicates that multiple parties locally train their DNN structures and jointly aggregate a standard model under the orchestration of a parameter server or many non-colluding servers interacting with each other. After a complicated cloud-based training, a trained model will be released for further deployment, so-called ``machine learning as a service'' (MLaaS). It suggests that MLaaS paradigm that utilizes the cloud infrastructure to generate models and works as an application programming interface (API) as a service, offering online prediction services to clients. 

In the new cloud-based neural network applications, privacy is a game between attacker and defender.
Many researchers have conducted excellent work to explore privacy attack and privacy defense in the new area. Fig.1 summarizes the taxonomy of ML privacy game in neural network and Fig.2 illustrates the general workflow and the platform of the cloud-based neural network.
\begin{figure*}[!thbp]
	\centering
	\includegraphics[width=.7\textwidth]{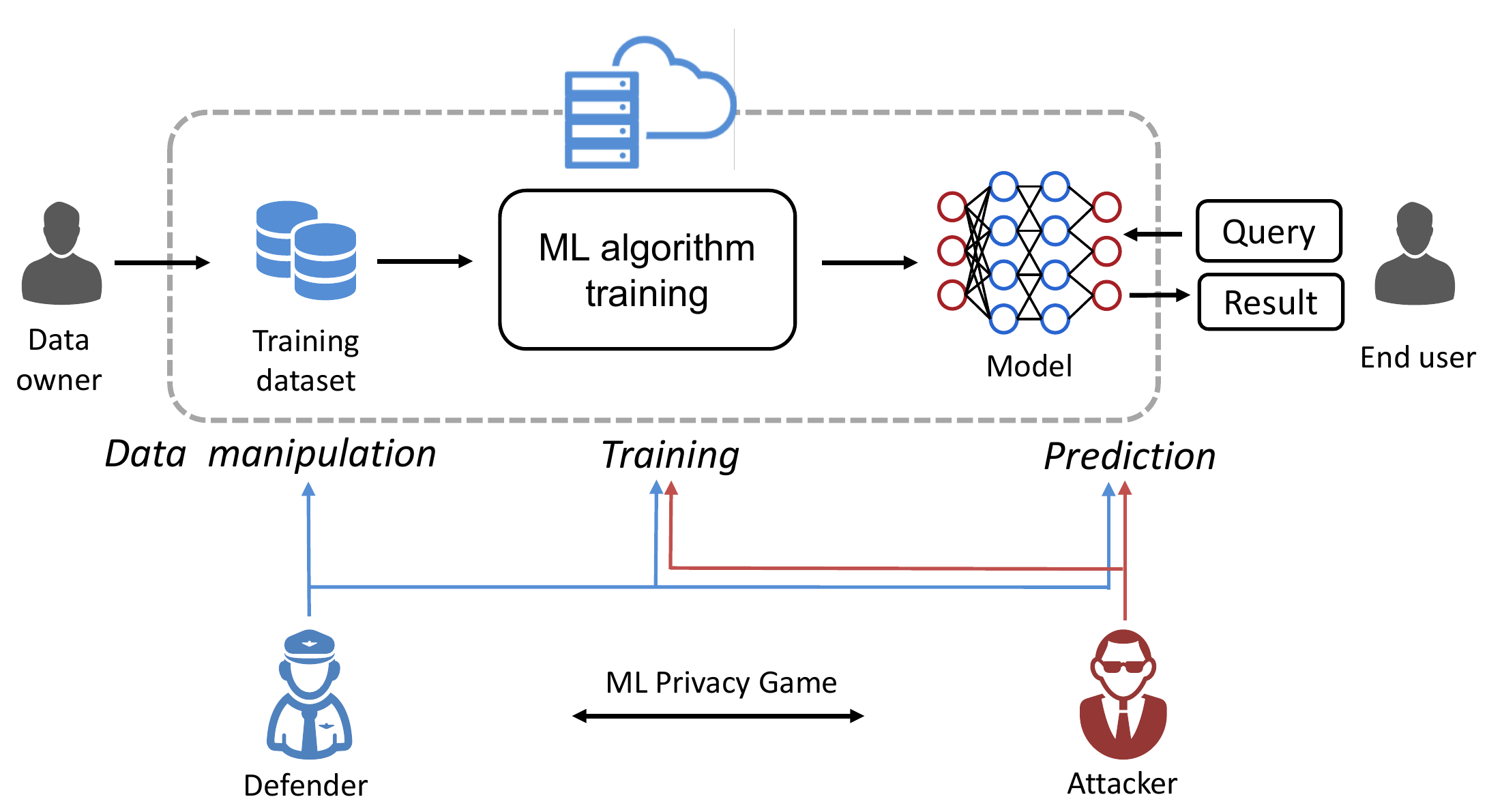}
	\caption{ML privacy game in cloud-based neural network service. Data from the user(s) are collected in ciphertext form and fed into a data warehouse for further training. Then, the primary training efforts focus on the cloud server-side, generating and publishing a machine intelligence model that can be deployed in a prediction service to serve future user queries.}
	\label{figure_system_model}
\end{figure*}
\subsection{Privacy attack}
There are four types of privacy attacks in our surveyed papers, \textit{i.e.,} property inference attack (PInf), model extraction attack (MExt), model inversion attack (MInv), and membership inference attack (MInf). 
In the training stage, the privacy leakage, including PInf and MInv mainly occur in a federated learning scenario, which refers to as the white-box attack.
PInf was firstly formulated by Ateniese et al. \cite{ateniese}. More precisely, given the intermediate results of the updating model, the main objective of property inference attack launched by insider participant is to establish a meta-classifier to infer some key statistical properties preserved by the other participants' training datasets. MInv was firstly proposed by Fredrikson et al. \cite{fredrikson1,fredrikson2} in pharmacogenetics privacy, which targets to acquire information that is related to the patient's genetic markers. There exist two adversarial model inversion scenarios, called data reconstruction and training class inference. For the first scenario, an insider attacker can reconstruct unknown data by observing the intermediate results of the updating model. In the second scenario, an attacker can produce records that look similar to samples from other victims.

In the prediction phase, a question here to pose is, whether it is possible to make the model oblivious. While it provides accurate predictions to end-users, users infer nothing about the target model's structure and parameters as well as protecting the query data and its results. Hence, considerable efforts are devoted to infringing on data privacy and protecting the target model against privacy inference attacks from two opposing perspectives.
For privacy leakage risks occurred in prediction phase, 
there are four main categories, including Model Extraction (MExt), Membership inference (MInf), Property inference (PInf), and Model inversion (MInv). In details, MExt means that an attacker with black-box or grey-box access but no prior knowledge of the parameters or structures of the models, aiming to duplicate the functionality of this model, \emph{i.e.}, stealing/copying the target model; MInf implies that given a data sample and black-box access to a target model, an attacker can decide whether it comes from the target classifier's training set. MInf can pose severe risks, \textit{e.g.,} if the target model is trained on the datasets aggregated from persons with cancer, given a piece of victim's data point with knowing it belongs to the classifier's training dataset, then the adversary can know the victim's health condition. Recently, MInf has been successfully explored in many various fields, for instance, mobility data \cite{pyrgelis} and biomedical data \cite{backes}; PInf aims to deduce some properties related to the training samples by training a  meta-classifier  to  determine whether  the  target  model  has  a  specific  property; the common strategy  for MInv is to design  a  training-based  attack  model. More concretely, an attacker only knowing the prediction values (even some truncated values)  aims to recover the target model's specific training samples.



\subsection{Privacy defense}
In general, cryptography and differential privacy (DP) are two most common approaches to achieve the goal of ``secure data manipulation'', \textit{i.e.,} protecting users data privacy. The working principle of cryptography is to adopt some cryptographic primitives, for instance, secure multi-party computation, homomorphic encryption, secret sharing and oblivious transfer, to transfer plaintext into ciphertext. So the subsequent entities perform ML training or inference on encrypted data. Unlike above, DP is utilized to add random noise to the original input data resulting in fuzzy outputs. Besides, adversarial machine learning can also be employed to ensure data privacy in secure data manipulation. To sum up, secure data manipulation aims to privately transferring the training data from data owner(s) to the cloud server provider who provides an integrated training platform and rich computing resources.

For privacy defense in the training phase, the approaches consist of four categories according to the defense strategies, namely differential privacy, adversarial machine learning, watermarking and cryptography. 
The first class aims to obfuscate model's intermediate parameters by adding random noise in the attempt to keep their training datasets private; the second class aims to introduce the min-max game into the training process to enhance the robustness of the model in an adversarial learning manner; the third class intends to embed the watermarks (some pre-defined pairs or acting as the regularizer ) during the training procedure, using them for model ownership verification; the last category works on this specific setting where two or more non-colluding servers collaboratively train a common classifier and thus this problem is formalized as a secure multi-party computation scenario naturally.

For privacy defense in the prediction phase, the goal is to ensure that the model owners learn nothing about the query data while end-users learn nothing about the target model. To be more specific, there are two defense approaches, \emph{i.e.}, cryptography and adversarial machine learning methods. Based on some cryptographic tools, one can encrypt both the target model and the query data, then to make predictive results oblivious. Hence, the target model and the query data are transferred into ciphertext forms so that predictions can be done in an oblivious way. Besides, instead of modifying the model's training procedure, in this phase, it is designed to add noise to the target models' outputs via adversarial machine learning to defend against membership inference attack \cite{jia1}. 

\section{Privacy attack strategy in cloud-based NN}
We review all the crucial papers published in the past five years on privacy attacks in a deep neural network. It will be elaborated methodically along the workflow proposed in Fig. 2, including the training and the prediction phases. First of all, we review the privacy attacks in a collaborative learning setting, where multiple participants jointly build a shared model with their datasets by training the model locally and iteratively updating model gradients globally. In this process, the adversary acts as a malicious participant who could deduce some confidential information from other benign participants. As a result, the training dataset privacy will be leaked, such as some properties and data distribution. When deploying the perfect trained model, predicting as a commodity, it also raises privacy concerns concerning the sensitive information, such as the training data, model parameters and query samples. Notably, in this phase, privacy attacks mainly include two categories, \textit{i.e.,} white- and black-box attack mode. In terms of white-box attack, an attacker knows the target model's architecture and enables to acquire the internal parameters. For the black-box attack, the adversary can only request the target classifier via prediction APIs and obtain its corresponding confidence values. 

Due to their complexity and inscrutability, most NN models suffer from inference attacks and have an incredibly high risk of leaking information concerning their training datasets, model structures and parameters.
In the following, we introduce four kinds of malicious attacks based on different information stealing strategies, \textit{i.e.,}  PInf, MExt, MInv, and MInf. To make the privacy attacks occurred in the deep neural network be illustrated more systematically and comprehensively, a highly generalized summary is presented in Table I. This summary will facilitate the research community to understand, follow, and improve future work in privacy attacks involved in deep neural network learning. Table I illustrates each work's ``category” 
(including ``PInf”, ``MExt”, ``MInv” and ``MInf”), ``Phase” (including training and prediction), ``Target privacy information” (including training datasets and target model),  and ``System model” (including  ``CL”  and  ``SL” ). The characters ``CL” and ``SL”   in Table I are short for the collaborative learning and the standalone learning, respectively.

\begin{table*}
\begin{center}
  \caption{Outline of Reviewed Paper Concerning Privacy Attacks.}
  \label{tab:commands}
  \begin{tabular}{cccccccc}
    \toprule
     \multirow{2}{*}{Paper, Year} &  \multirow{2}{*} {Category} & 	\multicolumn{2}{c}{Phase} & \multicolumn{2}{c}{Target information}  & \multicolumn{2}{c}{System model} \\
     \cline{3-8}
          & & Training &  Prediction & Training data & Model  & CL & SL \\
\midrule
     \cite{melis}, 2019 & PInf & \checkmark & &  \checkmark &  & \checkmark & \\
     \midrule
       \cite{ganju}, 2018 & PInf & & \checkmark &  \checkmark &   & & \checkmark \\
       \midrule
       
            \cite{chase2021property}, 2021 & PInf & & \checkmark &  \checkmark &   & & \checkmark \\
       \midrule
                    \cite{tramer}, 2016 & MExt & & \checkmark & & \checkmark &  & \checkmark  \\
\midrule        
             \cite{wang}, 2018 & MExt & & \checkmark & & \checkmark  &  & \checkmark  \\
 \midrule            
       \cite{Orekondy}, 2019 & MExt & & \checkmark & & \checkmark  &  & \checkmark  \\   
 \midrule      
 \cite{Atli}, 2019 & MExt & & \checkmark & & \checkmark  &  & \checkmark  \\           
  \midrule           
    \cite{Jagielski},   2020 & MExt & & \checkmark & & \checkmark  &  & \checkmark  \\       
\midrule             
      \cite{hitaj}, 2017  & MInv & \checkmark & & \checkmark &  & \checkmark & \\
  \midrule

        \cite{song}, 2017 & MInv & &  \checkmark & \checkmark &  &  & \checkmark  \\
\midrule           
             \cite{yang}, 2019 & MInv & &  \checkmark & \checkmark &  &  & \checkmark  \\
\midrule

             \cite{shokri1}, 2017  & MInf & & \checkmark & \checkmark & &  &  \checkmark \\
  \midrule         
             \cite{song1}, 2019 & MInf  & & \checkmark & \checkmark &  &  & \checkmark \\
  \midrule         
             \cite{salem}, 2019  & MInf & & \checkmark & \checkmark & &  & \checkmark  \\
             \midrule
              \cite{nasr1}, 2019  & MInf & & \checkmark &  \checkmark &  & \checkmark & \checkmark\\
    \bottomrule
  \end{tabular}
  \end{center}
\end{table*}

\subsection{Property Inference Attack}
Property inference attack takes full advantage of the intuition that ML models learned from similar datasets and algorithms will show similar functions. The similarity of these functions will reveal some common inherent patterns among these trained models. The ultimate goal of an attacker is to discern these patterns and infer confidential information about the training dataset which the model provider does not 
prefer to release, for instance, the certain fraction of the training data from a specific category, the environment where the model was produced, the presence of an exact data point.

\begin{table*}[!h]
   \begin{center}
  \caption{Comparison of Property Inference Attack}
  \label{tab:commands}
  \begin{tabular}{cccccc}
    \toprule
   Paper, Year & Dataset & Mode  & Prior knowledge & Target information & Techniques \\
\midrule
\multirow{3}{*}{\cite{melis}, 2019} & LFW, FaceScrub, PIPA, & \multirow{3}{*}{White-box}  & Global model's  & Victim participant's &  Training a \\
    &  CSI, FourSquare,   &   & updates & training data & meta-classifier\\
      &  Yelp-health, Yelp-author & & & &\\

       \midrule

\multirow{3}{*}{\cite{ganju}, 2018} & US Census Income,   & \multirow{3}{*}{White-box} & Model's parameters & \multirow{3}{*}{Model's training data}   &  Neuron sorting;\\
    & MNIST, CelebA, & & and architecture  &  & Set-based  \\
      & HPCs &  &  & & representation\\
             \midrule

     \multirow{3}{*}{\cite{chase2021property},  2021} & Census,   & \multirow{3}{*}{Black-box} & Access to conditional  & \multirow{3}{*}{Model’s training data} & \multirow{3}{*}{Poisoning attack} \\
    & Enron & &  distributions,  Training &     &  \\
      &  & &  algorithm, the features &    &  \\
 
    \bottomrule
  \end{tabular}
  \end{center}
\end{table*}


Collaborative deep learning has become an alternative to conventional ML framework, which allows multiple users to build a shared model via training and updating local model structures with their datasets. However, some unintended information leakage issues arise during the model parameter updates in collaborative learning. 
In the PInf settings, it falls into two types relying on whether affecting the global model training procedure, \textit{i.e.}, passive property inference and active property inference \cite{melis}. The main idea behind PInf was that an attacker enabled to use  the snapshots of the global classifier's updates to generate a collection of training samples on labels with/without a specific property. Then, it locally built a binary property classifier during collaborative learning. The different procedure with the active inference attack was that an adversary appended additional loss function to the primary one locally and submitted it into the original collaborative learning procedure, forcing the global model to study separable representations for the samples with/without the property. Thus, the gradients would be separable, making it feasible for the attacker to learn if the training set had such property. 
The effectiveness of the property inference attacks were evaluated on two datasets, \textit{i.e.},  FourSquare  \cite{yang1}, Yelp-health. The results showed that as the size of batch rises, an attacker saw more words per batch, and it outputted more false positives. Besides, there existed no false negatives; the reason was that any real test must be contained in at least one of the batch collected by the attacker. 
The following experimental results showed that 1) the gradients acquired during collaborative training revealed more information than the representation of each category; 2) due to the active attack mode, the shared classifier learned a better separate representation for samples with or without of property. 


As it is well known that  the conventional route of property inference attack is to train a meta-classifier to determine whether the target model has a specific property. To this end, the attacker generates a series of shadow models to imitate the target classifier's behaviours. Then, the shadow models' parameters are leveraged to learn the meta-classifier.  However, it is shown that this method performs not well when employed into DNNs due to their complexity and inscrutability. Sometimes, the Fully Connected Neural Networks (FCNNs) make the extracting this property information more challenging. Thus, two techniques were exploited to alleviate these challenges by observing that FCNNs were invariant thanks to the permutation equivalence  \cite{ganju}. The first method was to arrange the FCNN into a concise form by sorting neurons without changing the function that it presents. In this way, all permutation equivalents possessed the identical representations. Especially, the magnitude of each node's sum of all the weights was used as the metric for further sorting. Then, for the FCNN, the output of each layer can be represented as an ordered vector; the second method was to apply set-based representations for learning the meta-classifier. As the name implied that a neural network $f$ could be viewed as an ordered collection of fully connected layers, in which each layer could be illustrated as a battery of neurons.
Four real-world datasets (US Census Income, MNIST, Celebfaces Attributes and Hardware Performace Counters) were used to evaluate these two methods of property inference attack. The experimental results showed that the Baseline method did not perform well on the majority of experiments with accuracy from a range of 55\% to 77\%. Compared to the Baseline approach,  the sorting-based method always outperformed, while the set-based approach was performing even better with the attack accuracy ranging from 85\% to 100\%. It demonstrated that an effective property inference attack on FNN could be launched by exploiting the sorting-based and set-based methods. 

  Feed-forward neural networks have shown excellent performance in many complex task fields, prompting their accelerated promotion in all walks of life. However, it also brings some safety and security issues due to the lack of robustness. For example, a property inference attack, which could get the property information about the training datasets, will seriously affect the privacy of the model. Besides, a poisoning attack could decrease the accuracy of a well-trained model. A poisoned dataset will contain some malicious sample added by the attacker. 
	Recently, Chase et al. \cite{chase2021property} focused on new property inference scenarios in which the attacker would launch a poisoning attack to increase the privacy risk of the global model when played a role as a collaborative learning participant. By regarding the target model as a Bayes-optimal classifier, an attacker could use the distribution information of the training subset which had different properties to inference the properties of the original datasets. Specifically, the attacker also discussed the selecting strategies of poison samples and query samples.
To evaluate the performance of the methods, the authors run their experiments on two datasets: US Census Income Dataset (Census) and the Enron email dataset (Enron). The results showed that it did indeed succeed with a very high probability. It proved that a poisoning attack could be used by the adversary to increase the information leakage of trained models.

\textbf{Comparisons and Insights.} Sharing of the trained model becomes a growing trend with the wide adoption of deep learning. However, along with the convenience brought by model sharing, there also exists a risk that the model consumers may capture some other confidential properties of the training data. This type of attack is named  \textit{property inference attack}. 
Table II summarizes the current works related to the property inference attack, describing the different attack modes, prior knowledge, target information, techniques, and real experimental datasets. As we can see from Table II, property inference attacks often conducted in scenarios where a malicious adversary can participate in the training procedure or capture the parameters of the target model, \textit{i.e.}, white-box mode. Besides, current work also showed that poisoning attacks can boost property inference attack where occurred in the black-box setting.
That is to say, an adversary can only access the predictions of the target model under the condition that it injected the poison data into the training set.
The attack goal of an adversary is to imply a certain desirable property or extract some global statistics of the training samples from the trained model that the model developer does not intend to leak. The work in \cite{melis} focused on stealing property information on individuals, i.e., the victim participant. Whereas, both \cite{ganju} and \cite{chase2021property}  investigated inferring the global property of the training set from the trained model. 
For defense against property inference attacks on the global properties of the training set, cryptographic techniques, differential privacy mechanisms, and any hardware-assisted solutions may be invalid in mitigating this global property information leakage. The reason is that all of these methods  do not affect the global property of the training set through black-box access to the prediction model. Notice that the property being deduced does not have to be an explicit feature in training samples, nor does it need to be related to the training data labels.

Moreover, a general strategy for launching a property inference attack is to collect substantial training datasets and then use them to train a meta-classifier locally to determine whether the target classifier has a specific property or not.
For some uncommonly encountered cases, the techniques of equal treatment are introduced to align with the conventional route.

\subsection{Model Extraction Attack}

The machine learning model can be deemed as a kind of valuable intellectual property due to its expensive training process. Model extraction attack has emerged as a new privacy risk as to the popularity of machine learning as a service (MLaaS). The objective of an attacker is to duplicate the functionality of the target model, \textit{i.e.}, stealing the target model. Model extraction attacks are often conducted in which a malicious adversary is only allowed to access the model via API calls or gain parts of the model parameters (the training set, the ML procedure, 
or the target classifier's parameters), \textit{i.e.}, black- or grey-box mode. Furthermore, a general strategy for launching a model extraction attack is to mine the relationship between the input feature vectors and the predictions or observe the objective functions' performance in training data.


Given the black-box access to the ML classifier, an adversary with no prior knowledge aims to steal the target model $f$ through the MLaaS platform.
That is, this type of attack targets at duplicating a surrogate model $\hat{f}$ which closely approximates or even matches $f$. Different to the traditional machine learning scenarios, MLaaS systems take the feature vectors as inputs and provide the adversary with some information-rich query results, which include confidence values with predictions. Then, the adversary leverages these confidence values to perform model extraction attack. The critical components of model extraction attack targeting deep neural networks include query input designing, confidence value collection and attack using equation-solving method \cite{tramer}. 
To recover the parameters of the target classifier $f$, the equation-solving method was adopted to solve this hard  problem. As illustrated in \cite{tramer}, the API provided by MLaaS outputted class probabilities matching the input query sample $x$, which naturally formed a tuple ($x, f(x)$). It implied that these tuples could be viewed as many equations, and were pretty fit for the unknown model parameters. Fortunately, for large quantities of models' output class, these equation systems would be solved more efficiently, resulting in stealing the model parameters $f$ precisely.
For the experiments, they focused on deep neural networks which only contained one hidden layer and evaluated this attack on multi-class models. For a budget of 100$\cdot k$, where $k$ denoted the number of model parameters, it achieved $\textbf{\emph{R}}_{test}=99.16\%$ and $\textbf{\emph{R}}_{unif}=98.24\%$, utilizing 108200 requests each model on average.


Hyperparameter plays a vital role in machine learning and can be viewed as the global variable, such as the $K$ for $K$-NN, dropout rate \cite{srivastava}, and the mini-batch for DNNs. Significantly, given the training data, ML models trained with different hyperparameters differ in performance for testing. Besides, hyperparameters are often learned by consuming  extensive computing resources, and it appears to be achieved by exploiting proprietary algorithms that vary with diverse model learners. Thus, hyperparameters are regarded as valuable and confidential information. 
In \cite{wang}, the adversary launched an attack with a threat model where the training set, the ML procedure and the model parameters were known. This threat model was constructed on the fashionable cloud-based MLaaS platform, \textit{e.g.} Amazon Machine Learning \cite{amazon}, where an adversary might be an Internet client of the MLaaS platform. 
If the parameters were unaware in advance, an adversary could leverage the model parameters stealing attacks \cite{tramer} to acquire them first. Based on a thorough analysis of the general model training process, a significant observation showed that the values of model parameters were given with the  objective function goes down. It implied that the gradients of the objective function at the model parameters were all close to zero. Therefore, setting the gradients to \textbf{0} constituted a series of linear equations concerning hyperparameters.  Finally, by leveraging the linear least square approach \cite{montgomery}, the hyperparameters were ultimately determined.
In order to evaluate the validity of the hyperparameters stealing attack, 3-layer NNs for classification and regression algorithm with several real-world datasets were studied respectively. In the empirical experiments, relative estimate errors were applied, and the results demonstrated the high performance of hyperparameters stealing attack, with estimate error rates below 10\%.

\begin{table*}
 \begin{center}
  \caption{Comparison of Model Extraction Attack}
  \label{tab:commands}
  \begin{tabular}{cccccc}
    \toprule
   Paper, Year & Dataset & Mode  &  Prior knowledge   & Stealing information & Techniques \\
\midrule
\multirow{3}{*}{\cite{tramer}, 2016} & Blobs, 5-Class, Adult, & \multirow{3}{*}{Black-box} & \multirow{3}{*}{-}  & \multirow{3}{*}{ Model parameters}  &  Equation-solving;\\
    & Iris, Steak Survey,   & & & & Path-finding algorithm\\
      &  GSS Survey, Digits& & & &\\

\midrule
        
        \multirow{3}{*}{\cite{wang}, 2018} & Diabetes, GeoOrig, & \multirow{3}{*}{Grey-box} & Training set,  & \multirow{3}{*}{ Hyperparameters}   & Linear least \\
    & UJIIndoor, Iris,  & & ML algorithm,  & & square approach\\
      &  Madelon, Bank & &   parameters & & \\
  
\midrule
\multirow{4}{*}{\cite{Orekondy}, 2019} & Caltech256 \cite{Griffin}  & \multirow{4}{*}{Black-box} & \multirow{4}{*}{-}  & \multirow{4}{*}{ Functionality}  & \\
    & CUBS200 \cite{Wah} & & & & Training ``Knockoff" \\
      & Indoor67 \cite{Ariadna} & & & & classifier\\
       & Diabetic5 \cite{eyepacs} & & & &\\

             \midrule
              \multirow{3}{*}{\cite{Atli}, 2019} & Caltech, CUBS & \multirow{3}{*}{Black-box} & Training data  & \multirow{3}{*}{ Functionality}   & Given more realistic \\
    & Diabetic5, & & distribution;  &  &  assumptions\\
      & GTSRB, CIFAR 10  & & unlabeled data  &  &  \\

           \midrule
           \multirow{2}{*}  {\cite{Jagielski}, 2020} & \multirow{2}{*}  {MINST, CIFAR-10} & \multirow{2}{*}  {Black-box} & \multirow{2}{*}  {-}  &  Weight recovery;  & Learning-based strategy;  \\
& & & & Functionality &  Extraction model’s weights  \\
    \bottomrule
  \end{tabular}
  \end{center}
\end{table*}



As we mentioned before, a few model stealing attacks targeting different parts of the model, such as parameters \cite{tramer}, hyperparameters \cite{wang}, architecture \cite{Oh}, and decision boundaries \cite{Papernot1}, had been investigated. It laid the foundation to precisely duplicate the target model. Towards relaxing some additional assumptions, Orekondy et al. \cite{Orekondy}
focused on model functionality stealing and developed the first method to reproduce the victim model with high accuracy as well as under the constraint to minimize the access to the black-box mode. 
More specially, they formulated model  functionality stealing as two phases, including transfer set construction and training a replicate model, named knockoff. The first step was that using a random strategy and an adaptive strategy to sample publicly-available images from the distribution $P_{A}(X)$, and then fed the victim model with a set of random images and obtained the corresponding predictions to form a transfer dataset, \textit{i.e.}, input-prediction pairs. Then, the second step was to train knockoff, aiming at imitating the victim model on the transfer set via minimizing the standard loss function. Knockoff were evaluated on four dataset including Caltech256 \cite{Griffin}, CUBS200 \cite{Wah}, Indoor67 \cite{Ariadna}, and Diabetic5 \cite{eyepacs}, in terms of \emph{accuracy} and \emph{sample-efficiency}. Note that accuracy was depicted in two versions: absolute ($x\%$) or relative to black-box victim model $F_{v}$ ($x\times$). The extensive experimental results showed that all knockoff classifiers reproduced $0.92-1.05\times$ performance of the victim model $F_{v}$. Besides, it was also demonstrated that model functionality stealing worked well in real applications due to knockoffs' strong performance, only USD $\$30$ were required to query the victim model.



Following Knockoff Nets \cite{Orekondy}, Atli et al. \cite{Atli} investigated how much information related to the target model can be extracted by an adversary via a black-box prediction API in real-world ML scenario. 
They evaluated the Knockoff Nets against five complicated victim DNN models, demonstrating that it could actually extract surrogate models and then confirming its good performance.
Then, the adversary model in \cite{Atli} was revisited and evaluated under different experimental setups, revealing its several limitations: 1) the different architecture between the surrogate model and the victim model resulted in low performance of the proposed Knockoff Nets; 2) if the granularity of the victim model's prediction was reduced, \textit{i.e.}, only given the classification result (one prediction class) or some truncated results (top-$k$ predictions), the performance of the surrogate model would degrade. Hence, attack effectiveness of Knockoff net would be substantially degraded when it was implemented in real-world API predictions. To enhance the attack effectiveness, Atli et al. \cite{Atli} designed a more realistic attack model. In terms of transfer dataset construction, introducing more representations for underrepresented categories or deleting a number of training records with fewer confidence values might contribute to a more successful attack.
In this way, attack effectiveness could  be dramatically increased with these realistic assumptions, and all existing defense techniques were no longer applicable. In their extensive experimental evaluations, the performance of Knockoff nets was measured by training 5-layer GTSRB-5L and 9-layer CIFAR10-9L models, leveraging Adam optimizer with the learning rate of 0.001 and 0.0005 over 100 epochs and 200 epochs, respectively. The results showed that even if the adversary used a different architecture, Knockoff nets performed well on the condition that both the victim and attack models used pre-trained architectures \cite{Erhan}. The experimental results conducted on Caltech-RN34, CUBS-RN34, Diabetic5-RN34, GTSRB-RN34, and CIFAR-RN34  showed that the accuracy of model extraction attack degraded when the prediction API outputted the truncated vector or only predicted class labels.




Recently, Jagielski et al. \cite{Jagielski} argued that any learning-based strategy prevented the attack model from extracting an accurate victim model with high-fidelity through comprehensive and analytical evaluations. However, both of two objectives: accuracy and fidelity, were desirable. Accuracy measured the predictive ability of the extracted model when encountering unseen test set. Fidelity depicted the general agreement between the replicated and victim models on any input samples. However, a high-fidelity extraction resulted in a relatively low performance due to the imperfection of the victim model while an extracted model with high accuracy would not replicate the errors of the victim, leading to a low-fidelity extraction. To break through this bottleneck, Jagielski et al. \cite{Jagielski}  presented a hybrid strategy to mount the functionally-equivalent model extraction attacks, which was also capable of directly extracting the victim model's weights. Whereas, directly performing a fully-learning-based attack would leave many free valuables to be solved, leading to functionality-equivalent extraction to be infeasible. Hence, the authors combined the learning-based strategy and the weight recovery strategy together, while fixing many of the variables to values recovered from extraction attack and learning the reminder of the valuables. Also, the proposed attack had a high adversarial example transferability, meaning that adversarial examples generated from the extracted model could also fool the remote victim model.  The results showed that this hybrid strategy increased the fidelity of the extracted model compared to prior work. For the worst-case example (only with direct extraction), the extracted model with a 128-neuron network had $80\%$ fidelity, while the fidelity agreement jumped to $99.75\%$ when performing the learning-based recovery strategy.

\textbf{Comparisons and Insights.} 
Parameters, hyperparameters, and architectures are critical in machine learning models because different values and forms often contribute to models with a large difference in performance. The goal of the model extraction attack is to plagiarize the model function from the victim classifier through black-box or grey-box access to prediction API. The generalized methods constitute learning-based strategy \cite{Orekondy, Atli} and weight recovery attacks such as equation-solving, path-finding attack \cite{tramer}  and linear least square approach \cite{wang}. The state-of-the-art work \cite{Jagielski} focused on combining learning-based and weight direct recovery strategies together to achieve both two objectives: accuracy and fidelity. Table III summarizes five kinds of current model extraction attacks, demonstrating their comparisons and differences in terms of experimental datasets, access mode, prior knowledge, stealing information, and attack techniques.

We conclude that there maybe exist several benefits to launch a model extraction attack. 1) \emph{Enjoy the target model for free}: the model developer hosts the target model in the cloud-based server and offers black-box access to the Internet users by providing an API. By charging for each query to the API, the malicious adversary will attempt to steal this valuable target model and use it for free. As long as the cost for the attacker to steal the model is lower than the cost for retraining, it is motivating enough for the attacker to do that. 2)  \emph{Leaking the privacy information of the training set}: the training data privacy can be inferred by malicious attackers who pay to query the target model because the target model is trained and obtained from the training data. Hence, extracting the target model information or function and analyzing the stolen model result in the training data privacy leakage. 3) \emph{Bypassing security detection}: in an ever-growing number of application scenarios, ML models are used to detect malicious behaviors, such as spam filtering, malware detection, and network anomaly detection. After extracting the target model, an attacker can construct the corresponding adversarial examples according to some relevant knowledge to bypass the security detection model. To sum up, model extraction attack does serious harm to the model copyright protection, privacy of the training data as well as model security, and so on.

\subsection{Model Inversion Attack}

Model inversion attacks explore the fact that the trained classifier can memorize information regarding the training dataset. That is, the trained classifier or the model updated in a collaborative training procedure contains some inherent information related to the training set, posing the highly sensitive information of data providers at risks.
The objective of an attacker is to deduce some useful content about the target classifier's training data, other participants' training data in collaborative learning, or the query samples from the classifier's prediction vectors. Model inversion attacks are often implemented in scenarios where an attacker can join in the training procedure or capture the prediction values of the target classifier, \textit{i.e.}, white- or black-box mode. 


Different from the traditional centralized training manner,  collaborative deep learning is proposed to tackle the problem of direct access to the sensitive datasets stored on the cloud platform. It works in the mode of which all parties train their local models and  release a little fraction of the model parameters, aiming at keeping their training dataset privately. Moreover, the shared model parameters could also be blinded via DP, attempting to force the information extraction attack infeasible \cite{shokri}. However, the privacy leakage problem was not well dealt with by using the methods mentioned above. An attack against privacy-preserving collaborative deep learning was devised to steal sensitive information about any other participant in \cite{hitaj}. More specifically, the adversary leveraged the training process to privately learn a Generative Adversarial Network (GAN) for generating prototypical data points of victims, \textit{i.e.}, stealing the training samples that were meant to be private. 
As an inside attacker, the goal of the adversary was to induce the victim to release more information about class $a$ (unknown to an attacker) without disturbing the distributed learning procedure. More explicitly, the adversary injected some fake data points from class $a$, as class $c$ into the learning process deliberately. To improve the accuracy of the global model, the victim needed to work hard to distinguish class $a$ and class $c$ while revealing more useful content concerning class $a$ than originally needed. Meanwhile, an attacker locally trained a GAN model to mimic the class $a$ from the victim. The real training dataset and a fake training dataset (released by the generator) were used to train the discriminator. The output of the discriminator and a fake training dataset (sampled at randomly but classified as $a$) was fed into the generator for training, and so forth. Once the prediction label released from the discriminator was $a$, the attack was successful.
In the experiments, this attack was performed on two real-world datasets, MNIST \cite{lecun1} and AT\&T dataset of faces \cite{samaria}. Compared with the existing model inversion attacks, this GAN attack achieved a remarkable result with high accuracy \textit{i.e.}, above 97\%. 


Machine learning becomes a commodity where end-users seek to apply the third-party ML techniques to their datasets to offer  ML services. Leveraging online machine learning services can save the cost of hardware and software maintenance. Though, it may put the sensitive data at risk. In other words, without knowing the working mechanism of the training process, end users may reveal their training data to the malicious ML service providers.
The scenario they consider in \cite{song} was that a vicious ML service vendor offered the learning algorithm code to the end-user without observing the whole training process. After creating a predictive model, the 
attacker had access to the entire classifier (``called white-box”) or only to a prediction API (called ``black-box”). In the case of white-box attack, three encoding techniques were exploited to extract the training sets from the end user: 1) encoding sensitive information concerning the training set in the model parameters' least significant bits also could achieve a high-precision model, 2) forcing the parameters to be positively correlated with the useful content, 3) embedding the useful message into the signs of the parameters; in the case of black-box attack, an approach resembled data augmentation without modifying the training procedure was proposed. 
Five-layer Convolutional Neural Networks (CNN) \cite{lecun} conducted on  LFW for gender-based classification was used to estimate attack effectiveness in four methods, including the LSB encoding attack, the correlated value encoding attack, the sign encoding attack and the capacity abuse attack. All the experimental results showed that the test accuracy was no less than 85\%, confirming that adopting the third-party ML algorithm to learn models on the sensitive dataset was quite risky.

\begin{table*}
\begin{center}
  \caption{Comparison of Model Inversion Attack}
  \label{tab:commands}
  \begin{tabular}{cccccc}
    \toprule
   Paper, Year & Dataset & Mode  &  Prior knowledge   & Target information & Techniques \\
\midrule
\multirow{3}{*}{\cite{hitaj}, 2017} &  MNIST, AT\&T  & \multirow{3}{*}{White-box} & The model's structure; & Victim participant's  & \multirow{3}{*}{GAN}\\
    &  dataset of faces  & & The data labels of  & training data &\\
      & & & victim participants & & \\

       \midrule

              \multirow{3}{*}{\cite{song}, 2017} & CIFAR10, LFW, & Black-box \&  & - \&   & Target model's  & LSB encoding; \\
    & FaceScrub, 20 Newsgroups, & White-box & Model information  & training data &Correlated value encoding; \\
      & IMDB Movie Reviews   & &  &  & Sign encoding\\
  \midrule

              \multirow{3}{*}{\cite{yang}, 2019} & FaceScurb, CelebA, & \multirow{3}{*}{Black-box} & Additional set;  & Target model's &  \multirow{3}{*}{Training-based strategy}\\
    & CIFAR10, MNIST  &  & Partial prediction & training data & \\
      &  & &  values  &  & \\
    \bottomrule
  \end{tabular}
  \end{center}
\end{table*}



Whereas, model inversion attack in adversarial scenarios \cite{yang} was very different from the existing inversion settings. Significantly, an attacker had no knowledge concerning the training samples and thus, the previous training-based and optimization-based approaches could not be directly exploited. 
Yang et al. \cite{yang} designed a training-based attack strategy, where 
a second NN $G_{\theta}$ was learned to launch the model inversion attack. Firstly, the training dataset for building an inversion model $G_{\theta}$ were sampled from a more generic data distribution (namely, additional set) rather than the original training data distribution. The reason was that the auxiliary samples still retain general input features, sharing the standard semantic features with the original dataset. Then, a truncation method was exploited to train an inversion model $G_{\theta}$, which made the trained model aligned with the usually acquired prediction results, \textit{i.e.}, partial prediction values. In the actual training process, the outputs of model's prediction on auxiliary data points were truncated to the identical dimension ($m$) of the partial confidence values on the data of victim,  then the truncated result was fed into the inversion model  $G_{\theta}$ for training. If $m>1$, it outputted the desired reconstructed input data; if $m=1$, it could be reviewed as the training class inference as we mentioned before. Besides, when the attacker was the model producer of the target model, then it could alternatively train the inversion model with the primary learning task by adding reconstruction loss function, targeting improving the quality of $G_{\theta}$.
Four benchmark image recognition datasets, including  FaceScrub \cite{ng}, CelebA \cite{liu}, CIFAR10 \cite{krizhevsky} and MNIST \cite{lecun1}, were utilized to assessed the validity of the presented attack. And, the extensive  results demonstrated that the auxiliary samples and the truncation method contributed to the effectiveness of the model inversion attack. It further indicated that the plentiful message embedded in the classifier's outputs also could be sufficiently extracted even though an attacker only captured some constrained knowledge. Besides, joint training with the primary model resulted in a better performance of the inversion model at the expense of a bit lower classification accuracy.

\textbf{Comparisons and Insights.}
Model inversion attacks target inferring sensitive information about the training data or the query data. Given a target classifier and some prediction labels, the goal of an adversary is to recover the input data via its predictions.
There exist two adversarial inversion categories comprising of data reconstruction and training class inference. The adversary is required to reconstruct an unseen data point which is precisely the inversion of the target model or recovers a semantically meaningful data point for a given class of the target model.
Model inversion attacks can work in both black- and white-box attack scenarios.
An attacker can also be acted as ML service providers, inside malicious participants,  and malicious model consumers via API querying.  Table IV presents a comparison of several model inversion attacks in terms of experimental datasets, attack mode, prior knowledge, target information, and attack techniques.

The strategies for launching a model inversion attack depend on the specific scenario. In general, they are largely divided into two types of strategies. The first approach is named the \emph{optimization-based} method, which inverts a classifier by leveraging gradient-based optimization in the input samples space.  In this case, a model inversion attack is launched to infer the target classifier's training classes by creating a representative sample for a given class label. In other words, the inversion task is implemented by seeking the optimal data for the target class. However, the optimization-based approach is ineffective against complex convolutional neural networks, because the optimization objective of this method does not capture the semantics of input data space. The second method is called the \emph{learning-based} approach, which inverts a classifier by training a second model to act as the inverse of the target one. It targets regenerating images from their input features including activations for each layer. To this end, the adversary trains a second model by utilization the prediction vectors on the target model's training data.

\subsection{Membership Inference Attack}

 The target of membership inference attack (MInf) is to deduce whether a given sample has been part of the targeted model's training set. MInfs are launched in which a malicious adversary can participate in the training procedure or query the target model through API, \textit{i.e.}, white-box mode or black-box mode. Moreover, a generalized strategy for launching membership inference attack mainly consists of three primary phases, 1) collection of shadow model data set; 2) generation of the training set for the attack model; 3) training the attack model, as described in Fig. 3. 
 Phase I intends to emulates the original training dataset as much as possible to generate a shadow model dataset, by leveraging different adversary prior knowledge. In Phase II, the attacker utilizes the imitated dataset to learn shadow models to imitate the target model's behaviors in order to generate the attack model's training set further. Phase III intends to produce an attack classifier whose aim is to recognize the difference between the target model's performance on the training data points and its behaviors on the inputs that it unseen before. 


\begin{figure*}[!thbp]
	\centering
	\includegraphics[width=.6\textwidth]{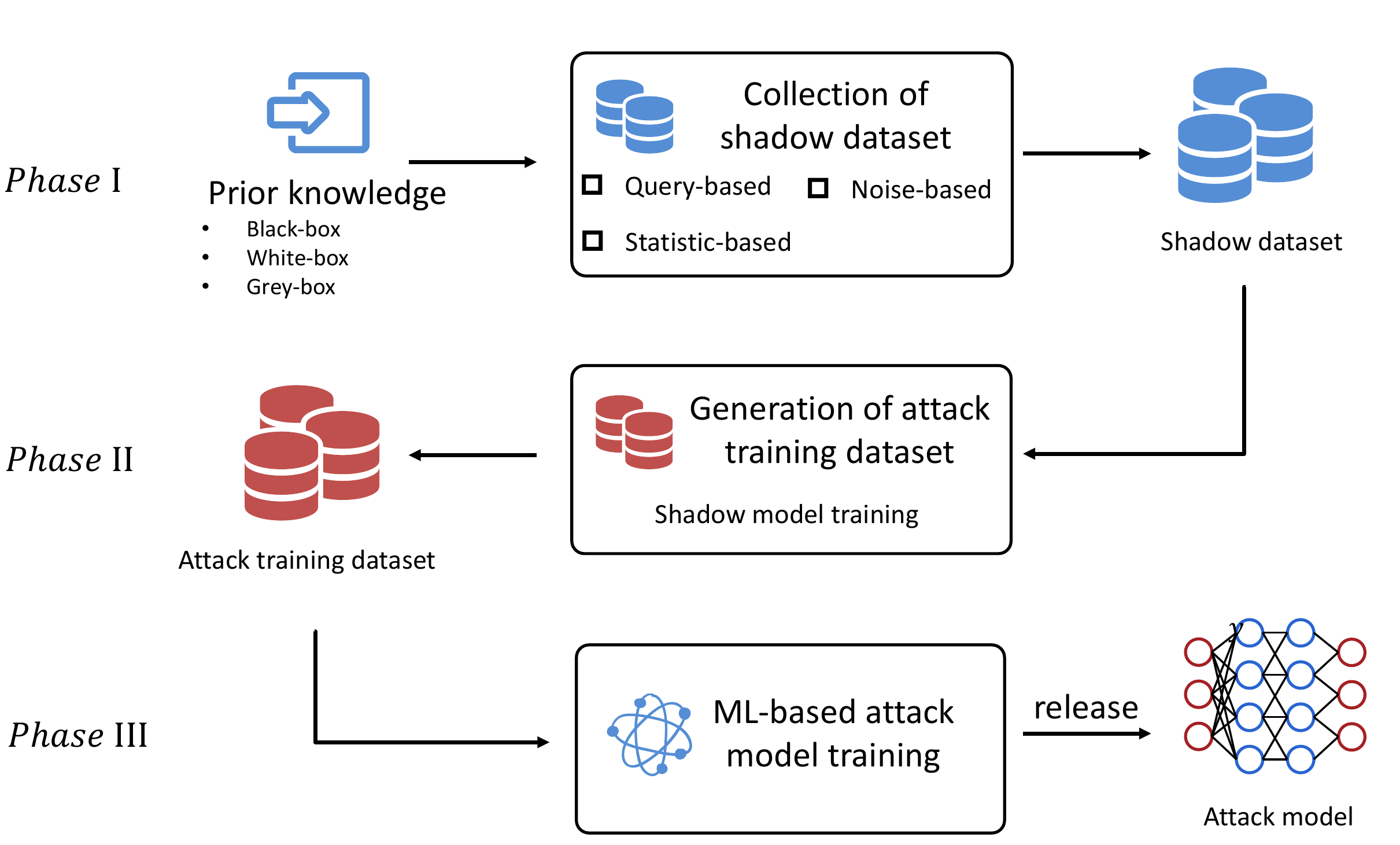}
	\caption{ The workflow of membership inference attack}
	\label{figure_system_model}
\end{figure*}

Given a sample and black-box access to the target classifier, MInfs \cite{shokri1} target determining whether this sample is part of the classifier's training set. Google and Amazon offer ``machine learning as a service”, making the prediction model available to the customers in a manner of black-box access (offered by providing API). Although the training sets and the model parameters are unknown to the model consumers, the model behaves differently on the inputs that it belongs to  the training dataset or not. According to the observation of such behaviors, an attacker can make adversarial use of ML and learn an inference classifier to identify and decide whether the input samples is part  of the target classifier's training set.
To achieve the goal of the membership inference attack, a shadow training technique \cite{shokri1} was leveraged, where a number of shadow models were created to imitate the target classifier's behaviors. Therefore, the training set and the corresponding ground truth about the membership were collected to constitute a corpus for training an inference model. Also, three effective methods were developed to producing training data for the shadow models, including black-box access-based synthesis, statistic-based synthesis and noisy-version access-based methods. Then, querying these shadow models with the training data and testing data, respectively, the prediction outputs were marked as ``in” or ``out”.  Naturally, an adversary made use of these records to build a binary classifier to infer the membership of a given record further. Remark that such offline attack was hard to be detected because the MLaaS system treated the malicious attacker as a benign user since he made queries online through API.
For its evaluation, several real-world datasets were used while three target models were built by Google Prediction API, Amazon ML and CNN, respectively. Besides, they evaluated this attack leveraging the standard metrics accuracy, precision and recall. Based on the experimental results, the accuracy of the constructed attack model was above the baseline 50\% among all experiments, while the precision was over 60\%, and the recall was close to 100\%. Therefore, it indicated that the membership inference attack model was successful with a higher precision of membership inference.


Whereas, the current proposed membership inference attack \cite{shokri1} was based on many assumptions, such as prior knowledge of the model structure, several shadow models used to mimic the target classifier's behaviors, and having the same data distribution as that of the target model. Nevertheless, all these fundamental assumptions were much stronger, significantly decreasing the feasibility of MInfs. Salem et al. \cite{salem} gradually relaxed all these assumptions aiming to demonstrate that such attacks can be widely applicable.
They proposed three types of adversaries in order to relax these assumptions gradually. Firstly, instead of training several shadow models to imitate the target classifier's behaviors, they only used one shadow model. Since that the shadow model was built through legitimate querying to MLaaS, adopting only one shadow model would significantly reduce the communication overhead of the membership inference attack. 
Secondly, they proposed a data transferring attack for shadow model training in order to prevent the attacker from learning the architectures and parameters of the target ML models. In this way, the shadow model was not utilized to simulate the behaviors of the target ML classifier, but only to learn the membership information of the training samples. 
Finally, the attack could be successful without using the shadow model, which implied that no training procedures were involved for mounting the membership inference attack.
They proposed a threshold-choosing method to differentiate the member and non-member data points by observing the statistical information, for instance, maximum or entropy leaked from the target classifier's posteriors.
Extensive experimental evaluations conducted on eight different datasets showed that with only one shadow model, the attacker could obtain a very close performance to that proposed by Shokri et al. \cite{shokri1}. 
Besides, the second adversary could achieve a slightly dropped performance compared with the first adversary. For the third adversary, this simple attack could still have valid inference ability over several data sets. In conclusion, MInf could be conducted in a much simpler and more powerful way, demonstrating the serious information leakage  of the ML classifiers. 

\begin{table*}
 \begin{center}
  \caption{Comparison of Membership Inference Attack}
  \label{tab:commands}
  \begin{tabular}{ccccc}
    \toprule
   Paper, Year & Dataset & Mode  &  Prior knowledge   & Techniques \\
\midrule
\multirow{3}{*}{\cite{shokri1}, 2017} & CIFAR, Purchases, Locations, & \multirow{3}{*}{Black-box} & Target model structure;     & \multirow{3}{*} {Multiple shadow  training}  \\
    &    Texas hospital stays,  & &  Training data distribution  &\\
      &  MNIST, UCI Adult &  & &\\

       \midrule

    \multirow{3}{*}{\cite{song1}, 2019} & Yale Face,  & \multirow{3}{*}{Black-box} & \multirow{3}{*}{Perturbation constraint $\mathscr{B}_{\epsilon}$} &    Confidence-thresholding  \\
    & Fashion-MNIST,  & &     &  inference strategy\\
      & CIFAR10  & &   &    \\
  \midrule

              \multirow{5}{*}{\cite{salem}, 2019} & Adult, CIFAR10, CIFAR100, & \multirow{5}{*}{Black-box} &      &\\
    & Face, Location, MNIST,  & & The same data distribution    &  One shadow training; \\
      &News, Purchase2, Purchase10,   & & - &   Data transferring attack; \\
        &Purchase20, Purchase50,    & & -  &   Threshold-choosing method\\
          &Purchase100   & &     & \\
      \midrule

              \multirow{3}{*}{ \cite{nasr1}, 2019} & CIFAR100,  & \multirow{3}{*}{White-box} &  Target model structure;     &  \multirow{3}{*}{White-box inference attack}\\
    & Purchase 100,  & &   a subset of training set  & \\
      & Texas100  & &     & \\
    \bottomrule
  \end{tabular}
  \end{center}
\end{table*}

 In particular, MInfs against the existing defense measures succeed in mitigating the evasion attack \cite{song1}. The intuition behind the membership inference attack in this setting was that adversarial defense method which resisted the evasion attack impacted the target model's precision boundaries. In other words, these defense methods worked through keeping the model predictions still for a small domain around per input data, increasing the decision-making power of records in target classifier's training set. Hence, the training data records had a more significant impact on robust classifiers, making these classifiers 
were more likely to suffer MInfs.
In order to measure the MInf on the robust adversarial models, two novel inference approaches were developed by leveraging the predictions of adversarial examples (AEs) and verified worst-case. The first approach adapted the PGD (Projected Gradient Descent) attack \cite{Madry} approach to seek AEs by iteratively minimizing the targeted loss function. For each class label, the attacker trained a binary membership inference classifier. 
A fraction of the training data and the query data were chosen and then used to generate targeted AEs. These AEs were fed into training the membership inference classifier. Then, the remaining training and test points were used to launch membership inference attacks. Different from the above heuristic PGD approach used to produce AEs, they utilized verification techniques utilized by the verifiable defended models \cite{gowal, mirman, wong} to obtain the input's worst-case predictions. Then, the worst-case confidence value of the input was leveraged to deduce its member status. 
The experimental results showed that compared to the undefended approach, adversarial defense method could definitely rise the risk of MInf. To be more specific, compared to the conventional training process, using the adversarial defenses to acquire the robust model increased the advantage of membership inference by up to 4.5$\times$.


Recent works had researched MInfs against black-box target classifiers, and these attacks had been proved to be robust against well-generalized DNNs \cite{shokri, yeom}. Nevertheless, directly exploiting the black-box membership inference attack methods into the scenarios where an adversary could observe or participant its training process was not infeasible. 
Nasr et al. \cite{nasr1} proposed a novel attack model, which exploited the white-box membership inference attack setting against deep neural networks, aiming at tracing the training data samples. In this architecture, it discussed different attack observations, different training algorithms (stand-alone and federated learning), different modes of the attack (passive and active), and different assumptions of adversary knowledge (supervised learning and unsupervised learning). This architecture constituted two main components: attack features extraction and attack model construction. The attacker run the target model and took the data sample $x$ as input. Then, it computed all the hidden layers $h_{i}(x)$, the classification result $f(x)$, the loss function $L(f(x),y; \textbf{W})$, and the gradients for each layer $\frac{\partial L}{\partial\textbf{W}_{i}}$. Then these computations as well as the true label $y$ were used to train the attack inference classifier. In terms of the attack model, it was comprised of convolutional neural network (CNN) as well as the fully connected network (FCN), followed by a fully connected encoder. For supervised learning, the single value of encoder's output  characterizes an encoding of the membership knowledge,  \textit{i.e.} $Pr(x,y) \in D$. For unsupervised learning, a decoder was deployed to reproduce the key features of the attack model. Then, the input of the target model was grounded into two clusters,  \textit{i.e.}, member and non-member.
To evaluate the validity of the proposed white-box MInf, they conducted the experiments on CIFAR-100. The extensive experimental results showed that the best model, \emph{i.e.,} DenseNet model with $82\%$ classification accuracy was not brittle to black-box attacks with a $54.5\%$ inference attack accuracy. However, it was vulnerable to the designed white-box inference attack with the accuracy of $74.3\%$. It 
demonstrated that the even well-generated classifiers might reveal a large quantities of information concerning training dataset.

\textbf{Comparisons and Insights.}
Membership inference attacks are mainly performed by using two techniques: shadow model training \cite{shokri1, salem} and threshold-choosing method \cite{song1}. The first approach is to imitate the target classifier's behaviors by introducing shadow model training and then further develop an attack model to distinguish the member status of an individual sample. The second method is to exploit some confidence-threshold and statistics-based methods to seek a proper threshold to decide whether a given data sample is part of the target classifier's training dataset. Table V presents five current membership inference attack works from the respect of the experimental dataset, access mode, prior knowledge, and techniques. It is not difficult to find out from Table V, a majority of membership inference attacks are in the black-box scenario \cite{shokri1, song1, salem}. That is to say, the attacker can only observe the model predictions. However, extending the known black-box membership inference attack mechanism directly to the white-box setting (by analyzing the output of the activation function) does not work well \cite{nasr1}, because the generalization capability of the activation functions is much faster than that in the output layer. Moreover, the black-box attack may be not effective for DNNs with good performance on generalization. Therefore, 
The current work  \cite{nasr1} leveraged the privacy vulnerability of the SGD algorithm used to train deep neural networks, and designed a novel white-box membership inference algorithm. The ultimate goal pursued by the privacy attack community is to gradually relax attack assumptions and make the designed attacks applicable for far more general attack settings.

We argue that one of the main reasons for a member inference attack is that the attacker can accurately infer whether a piece of data is in the training set of the target model. Overfitting means that the prediction ability of the model for the data from the training set is better than that for the test data, i.e. the non-training set.
 Therefore, overfitting models are more vulnerable to membership inference attacks. There are two explanations for model overfitting. The first one is that the target model is trained inadequately or excessively, which makes the trained model has a poor performance on the test data. The second reason may be that the training set is unrepresentative. If there is a significant difference between the distribution of the training set and the test set, the target model learned from the training set will show its diverse performance when an encounter with the test samples.

\section{Privacy defense strategy in cloud-based NN}

When faced with privacy attacks on sensitive information, tremendous efforts are devoted to the design of privacy-preserving machine learning. As depicted in the workflow of deep neural network learning in Fig. 2, these privacy defense measures target at preventing the training dataset's  sensitive knowledge, model parameters and query samples from being revealed. More importantly, the privacy defense strategies mainly focus on three phases as depicted above. First of all, the required massive data collection for further training tasks induces privacy concerns with respect to the training data. Through adding noise or blinding the input data, the training dataset can be transformed into blinding datasets and fed into the cloud server which is equipped with tremendous computation resources. It is well known that a large quantities of data collected from different sources contributes to highly-accurate deep neural networks.
Thus, in the training phase, efficient and secure protocols for multi-party machine learning tasks are required to develop and deploy afterwards. Especially, more than two non-colluding servers are introduced to improve the training efficiency. Hence, a breakthrough in secure two- or multi-party computation directly leads to a great improvement of privacy-preserving multi-party training algorithms. Then, when a perfect target model is released and deployed, it provides the Internet users with the prediction service. However, a large quantities of information concerning the training dataset, model parameters may be revealed through inferring the prediction results. As a result, the machine learning and security communities focus on taking effective measures relying on various defense strategies to defend against malicious confidential information stealing attacks. 

In particular, we classify these defense methods into four categories based on different strategies, \textit{i.e.}, differential privacy, adversarial machine learning, watermarking techniques, and cryptographic techniques. To make the privacy defense occurred in deep neural networks more systematically and comprehensively illustrated, a highly generalized summary is presented in Table VI. which is comprised of ``Phase'', ``Target privacy information,'' ``Defense strategy'' and ``System model''. In particular, the characters ``DM", ``T" and ``P" in line 2 refer to  data manipulation, training and prediction respectively; ``TD'', ``MP'' and ``QD'' are short for training data, model parameters and query data respectively; the terms ``DP'', ``AML'', ``WT'' and ``CT''  refer to differential privacy, adversarial machine learning, watermarking techniques and cryptographic techniques; the terms``CT'', ``FL'' and ``SL'' and ``Mul-S'' are short for collaborative learning, federated learning, standalone learning and multi-server learning scenario, respectively.

\begin{table*}
\begin{center}
  \caption{Outline of Reviewed Paper Concerning Privacy defenses}
  \label{tab:commands}
  \begin{tabular}{c c c  c c c c c c c c c c c}
    \toprule
     \multirow{2}{*}{Paper, Year} &  \multicolumn{3}{c}{Phase}  & 	\multicolumn{3}{c}{Target information} & \multicolumn{4}{c}{Defense strategy}  & \multicolumn{3}{c}{System model} \\
     \cline{2-14} 
          & DM & T& P & TD & MP  & QD & DP & AML & WT & CT & FL & SL  & Mul-S \\
          \midrule
          \cite{shokri},  2015  &  &\checkmark & & \checkmark & &   & \checkmark & & &  &  \checkmark & & \\ 
          \midrule
      \cite{abadi} ,   2016  &  & \checkmark & & \checkmark & &   & \checkmark & & &  &   & \checkmark & \\ 
    \midrule
    \cite{chen2020gs},   2020  &  & \checkmark & & \checkmark & &   & \checkmark & & &  &   & \checkmark & \\ 
    \midrule
       \cite{jia2}, 2018  & \checkmark & & & \checkmark & & & & \checkmark & & & & \checkmark & \\
       \midrule
        \cite{nasr}, 2018  & & &\checkmark & \checkmark & &  & & \checkmark & & & & \checkmark & \\
        \midrule
               \cite{jia1}, 2019  & & & \checkmark & \checkmark & &  & & \checkmark & & & & \checkmark & \\
               \midrule
            
\cite{nagai}, 2018 & & \checkmark &  & &  \checkmark &  & & & \checkmark & & & \checkmark & \\
\midrule
         \cite{zhang7}, 2018 & & \checkmark &  & &  \checkmark &    & & & \checkmark & & & \checkmark & \\
  \midrule       
      \cite{jia4}, 2020 & & \checkmark &  & &  \checkmark &  & & & \checkmark & & & \checkmark & \\
      \midrule
         \cite{szyller}, 2019 & &  & \checkmark & &  \checkmark &    & & & \checkmark & & & \checkmark & \\   
         \midrule
   \cite{bonawitz}, 2017  & \checkmark & & & \checkmark & &   & & & & \checkmark &  \checkmark & & \\         
      \midrule
        \cite{mohassel}, 2017  & & \checkmark & \checkmark & \checkmark & \checkmark & \checkmark  & & & & \checkmark & & & 2 \\
          \midrule
          \cite{mohassel1}, 2018   & &  \checkmark & \checkmark & \checkmark & \checkmark & \checkmark  & &&& \checkmark & & & 3 \\
\midrule
       \cite{agrawal}, 2019 & & \checkmark  & \checkmark & \checkmark & \checkmark & \checkmark & & & & \checkmark & & & 2 \\
       \midrule
 \cite{liu1}, 2017 & & & \checkmark &  & \checkmark & \checkmark  & & & &\checkmark & & \checkmark & \\
 \midrule         
             \cite{jiang}, 2018 & & & \checkmark & & \checkmark & \checkmark & & & &\checkmark & & \checkmark &\\
  \midrule     
            \cite{chen}, 2019 & & &\checkmark &  & \checkmark & \checkmark  & &&&\checkmark &  & \checkmark & \\

    \bottomrule
  \end{tabular}
  \end{center}
\end{table*}

\subsection{Differential Privacy}

Advanced differential privacy mechanisms \cite{abadi, bassily, chen2020gs} are combined with deep neural networks in order to protect training dataset privacy and model privacy almost without affecting its utility. It is executed in the training phase for two scenarios: 1) the customized noises are added into the shared gradients in collaborative learning to protect individual participants' private information or 2) noises are added into the local calculated gradients/objective function to develop a robust classifier against a variety of privacy attacks. The main goal for differential privacy machine learning is to seek a better trade-off of model utility and model privacy, \textit{i.e.}, an acceptable accuracy loss along with strong privacy guarantee.

Massive data collected and trained collaboratively contribute to higher accuracy of the deep learning models. However, it also poses the privacy of participants at risks. Although there are many privacy protection measures, for instance, secure multi-party computation  and homomorphic encryption, can be leveraged into privacy-preserving deep learning, low efficiency is the main bottleneck. Therefore, designing and implementing an efficient privacy-preserving deep learning algorithm becomes a key challenge in real-world applications. By exploiting the fact that optimization algorithms based on stochastic gradient descent can be operated in parallel, Shokri et al. \cite{shokri} presented an efficient privacy-preserving deep learning scheme. This method allowed participants to execute local deep learning algorithm independently on their private dataset and jointly trained a global model, sharing only a tiny subset of their model parameters. As a result, it could perform well on both model utility and model privacy.
However, sharing only a small part of model parameters still resulted in privacy leakage of the training dataset. Hence, in order to minimize this information leakage, Shokri et al. \cite{shokri} further introduced differential privacy into parameters updates by leveraging the sparse vector technique. 
Significantly, based on a consistent differential private mechanism, they randomly chose a tiny fraction of gradients whose values were above the threshold, and then shared the blinding values of the selected gradients. 
The influence of diverse values of the DP parameter $\epsilon$, the number of participants $N$ and the fraction of shared gradients $\theta_{\mu}$ were evaluated on the model accuracy. The bound $\gamma$ and the threshold $\tau$ were set to $0.001$ and $0.0001$, respectively. As expected, the experiment results showed that stronger differential privacy guarantees (smaller $\epsilon$ values) led to lower accuracy. When many participants shared a significant part of gradients, the prediction accuracy of the proposed deep learning approach outperformed the standalone learning.



Very recently, Abadi et al. \cite{abadi} combined DNN with advanced privacy-preserving mechanisms and further designed an efficient differentially private neural network learning model. Instead of adding conservative noise to the final model parameters, it targeted to control the impact of training samples throughout the learning procedure. More concretely, they focused on the SGD algorithm, which is the dominant optimization algorithm applied in NNs. At each round of iterative training, it randomly selected a subset of data records and computed the corresponding gradients of the loss function. Then, it  clipped the $l_{2}$ norm of each gradient, and computed the average and perturbed this value by adding noise with differential privacy mechanism. In the following, it took a fixed step following the opposite direction of the calculated noisy gradient. Finally, depending on the information preserved by the privacy accountant, the privacy loss of the mechanism needed to be computed when releasing a well-performed model. 
The proposed deep learning with differential privacy scheme was evaluated on two standard image recognition datasets, \textit{i.e.}, MNIST and CIFAR. The extensive experimental results demonstrated that deep neural network could achieve privacy protection at the modest cost of training efficiency, software complexity and model accuracy. 

To make full use of the dividends brought by the huge data scale, a lot of research topics about “collaborative learning” have emerged in the machine learning community, which could help to train models with data sets from several data owns. However, “data sharing” in some sensitive domains is risky and illegal.
Recently, several defense techniques for privacy-preserving collaborative learning techniques have emerged.
	Chen et al. \cite{chen2020gs} proposed GS-WGAN (Gradient-sanitized Wasserstein Generative Adversarial Networks), which could share a sanitized form of sensitive data with rigorous privacy guarantees. In contrast to prior work, this work could carry out a more fine-grained sanitization of sensitive information, as a sequence, the generator trained by this method generated more realistic samples. Specifically, due to the two limitations of the traditional DP-SGD algorithm, which would seriously affect the usability of the model, this paper used the chain rule to limit the scope of differential privacy noise to a subset of generator parameters in GAN. In this way, the privacy-preserving GAN could generate more informative samples. To verify the effectiveness of the method in the paper, the authors used MNIST and Fashion-MNIST datasets to test the privacy and utility (which include sample quality and usefulness for downstream tasks) of the scheme. The extensive evaluation showed that the sensitive datasets could be effectively distilled to sanitized forms with contain less sensitive information and hardly affected the training of the model.

\begin{table*}[t]
\centering
  \caption{Comparisons of Defense Strategies Based on Differential Privacy}
  \label{tab:commands}
  \begin{tabular}{cccccc}
    \toprule
   Paper, Year & Dataset & Model &  Target protected information    & Techniques \\
\midrule
\cite{shokri},  2015  &  MNIST, SVHN & MLP, CNN  & Gradients    & Distributed selective SGD \\

  \midrule

              \cite{abadi},   2016  & MNIST, CIFAR-10 & NN, CNN  &  Gradients      & Differentially private SGD \\
  \midrule

              \cite{chen2020gs},   2020  & MNIST, Fashion-MNIST & MLP, CNN,  11 scikit-learn classifiers & Part of the gradients   &  Selectively applying DP SGD\\
    \bottomrule
  \end{tabular}
\end{table*}

\textbf{Comparisons and Insights.}
Due to its benefits in low calculation overhead and comparatively well privacy guarantee, differential privacy mechanism is often integrated into conventional machine learning algorithms to protect individual participants' confidential information in the training dataset.  Table VII summarized several current privacy-preserving machine learning with DP mechanism from the respect of experimental datasets, training model, target protected information, and techniques. As we can see from Table VII, the main information that the model developer wants to preserve is the gradient information. The essential information that needs to be protected is the training dataset because the gradients are acquired from the training data samples.

Specifically, privacy-preserving machine learning approaches with differential privacy mainly include centralized learning and federated learning scenarios. We remark that applying differential privacy into deep learning models is one of the active protection mechanisms, which is taken by the model developer itself during the training procedure.
To achieve differential privacy with a modest utility loss, three noise-adding mechanisms are explored based on diverse requirements of specific scenarios, including objective function, internal parameters (such as gradients), and predictive confidence vectors. The ultimate goal of privacy defense measures combined with DP is to seek a better trade-off between model privacy and model utility. Additional improvements in accuracy and less privacy loss need to be obtained in future research.

\subsection{Adversarial Machine Learning}

Adversarial machine learning can be utilized to build privacy-preserving ML mechanisms by leveraging machine learning algorithms from an opponent perspective. The key idea behind this mechanism is that defenders can be assumed as  adversaries, so that they learn the behaviors of malicious attacker for better defense mechanisms from the perspective of privacy protection.

Attribute inference attack targets users' personal attributes, such as location, salary, hobbies and political view in various applications ranging from recommendation systems \cite{Otterbacher, Weinsberg}, to social network \cite{Zheleva, jia3}, to mobile platform \cite{Michalevsky, Narain}. It is launched by an adversary by leveraging an ML model to deduce a victim's private data whether it contains a particular attribute. 
Depending on the relationship between public data and attributes, the existing defense strategies mainly focus on making use of game theory or heuristics. Nevertheless, it is demonstrated that these defense strategies are not feasible in practice. Recently, Jia et al. \cite{jia2} proposed AttriGuard, a feasible defense strategy against attribute inference attacks, targeting decreasing utility loss as well as improving computation efficiency.
In particular, considering that the objective of AttriGuard was to prevent a user’s attribute from being leaked, it worked in two phases. In terms of the attribute's each value, the Phase I intended to seek a minimal noise such that it was added to the user’s data and the adversary's model was likely to predict the attribute value for the user. In addition, this noise was found by leveraging the advanced evasion attacks. For the Phase II, based on a particular probability distribution, one attribute value was chosen and the corresponding noise crafted in the Phase I was added to the user's public data vector. Finding the probability distribution can be solved as a constrained convex optimization problem. The proposed AttriGuard was compared with existing defenses by leveraging the same data sets from \cite{gong}. Three findings were concluded from the extensive experiments: 1) the adapted evasion attack used in the Phase I had a better performance than existing ones; 2) the proposed AttriGuard was shown that it can effectively defend against attribute inference attacks. For example, for some defense-unaware attribute inference attacks, the inference accuracy of attack classifier was decreased by $75\%$ when tampering at the top-4 rating scores on average; 3) AttriGuard required a tiny noise to users' data than the state-of-the-art defense strategies while decreasing the adversary’s inference accuracy by the same order of magnitude.

\begin{table*}
\centering
  \caption{Comparisons of Defense Strategies Based on Adversarial Machine Learning}
  \label{tab:commands}
  \begin{tabular}{cccccc}
    \toprule
   Paper, Year & Dataset & Model & Target information    & Techniques \\
\midrule
   \multirow{2}{*}{\cite{jia2}, 2018} &   \multirow{2}{*}{Google+ dataset} & Multi-class logistic regression,  &  \multirow{2}{*}{Attribute privacy}   &  \multirow{2}{*}{AttriGuard} \\
      &    & Neural network &    & \\
      
      \midrule

              \multirow{3}{*}{\cite{jia1}, 2019} &Location, & 6-layer FNN, & \multirow{3}{*}{Membership privacy}  & \multirow{3}{*}{MemGuard} \\
    & Texas100 , &  6-layer FNN, &    & \\
      &  CH-MNIST  & CNN &     & \\
 \midrule

              \multirow{3}{*}{\cite{nasr}, 2018} & CIFAR100, &  Alexnet, DenseNet & \multirow{3}{*}{Membership privacy}    &  \multirow{3}{*}{Min-max game}\\
    & Purchase100,   &  4-layer FNN&     & \\
      & Texas100   & 5-layer FNN &    & \\

    \bottomrule
  \end{tabular}
\end{table*}

MInf aims to deduce whether a given sample belongs to a target classifier's training set.
Most of the existing membership privacy defense approaches concentrate on leveraging differential privacy to regularize the learning procedure of the target classifier. However, these defense mechanisms have two crucial flaws: 1) for the confidence values, there is no guarantee for its formal utility-loss; 2) they may achieve an unsatisfied trade-off between privacy and utility.
Jia et al. \cite{jia1} presented a defense method named MemGuard against black-box membership inference attack through crafted adversarial examples. Concretely,  MemGuard appended noise to per prediction confidence values instead of modifying the training procedure of the target model. The fundamental intuition was that the attack classifier used by an adversary to determine the target sample's membership is vulnerable to the crafted adversarial examples. Then, MemGuard intended to transfer each  prediction into an adversary example that can misguide the adversary’s model. It worked in two phases: 1) seeking a crafted noise vector which transferred a benign confidence vector into an adversary example and misled the attack's classifier to infer the target sample's membership status; 2) the carefully-crafted noise vector was then added to the prediction with a certain probability, which was chosen to meet the requirement of a specific utility-loss budget. 
The attack classifier inferred membership status if and only if a value generated by the neuron in the output layer was larger than 0.5. Besides, the attack model was trained for 400 iterations with a learning rate 0.01 using the SGD algorithm and decreased the learning rate by 0.1 at the following 300 iterations. Besides, MemGuard was the first attempt to demonstrate that adversarial examples could be exploited as defense strategy against MInfs.

For MInf, as mentioned before, Nasr et al. \cite{nasr} focused on designing a rigorous membership privacy mechanism against the black-box attack with the minimum utility loss. The design objective  was to decrease the loss function and the probability of a successful membership inference attack launched by an adversary. Therefore, this optimization problem was conducted as a min-max privacy game between the inference attack and the corresponding defense mechanism, which was similar to other scenarios in \cite{alvim, hsu, shokri2}. They trained the classifier in an adversarial manner \cite{dumoulin, kozinski, miyato}, where the gain of the inference attack was treated as a regularizer and added to the original training loss function, then it was minimized along with the classification loss. The authors sought the trade-off between membership privacy and classification accuracy to further achieve the equilibrium point by tuning the regularization parameter. In this manner, where the loss function for achieving membership privacy was decreased, the most decisive MInf against the target classifier equaled to a random guess, indicating the maximum member status privacy.
The results showed that the proposed scheme enabled to achieve $76.5\%$ testing accuracy as well as $51.8\%$  membership inference accuracy on the Purchase100 dataset \cite{shokri1}. Using a standard L2-norm regularizer, by contrast, might present the same relative level of privacy (against the identical attack) but with a pretty low prediction accuracy $32.1\%$. Compared to a regular non-privacy-preserving model, the classification accuracy  dropped $1.1\%$ and $3\%$, respectively, for the CIFAR-100 performed with Densenet as well as architectures. In terms of prediction accuracy, it  dropped  from $67.6\%$ to $51.6\%$ and from $63\%$ to $51\%$ for the Purchase100 and Texas100 datasets, respectively.


\textbf{Comparisons and Insights.}
Defense mechanisms combined with adversarial ML can be deployed into data manipulation, training phase, and prediction phase, respectively. Adversarial machine learning used as a defense measure leverages the idea of considering the potential adversary. When designing a machine learning algorithm, it assumes in advance that there is a malicious attacker who may pose a privacy threat. Table VIII summarizes the comparison of defense strategies based on adversarial machine learning techniques from the respect of experimental datasets and models, the target protected information, and techniques. Almost without exception, the privacy defender utilizes reverse thinking and aims to protect the training data privacy.

The key idea of AttriGuard \cite{jia2} was to find a proper noise and added them into the pubic data to defense against attribute inference attacks. What's similar was that  MemGuard \cite{jia1} exploited adversarial training to mimic evasion attacks and generated adversarial examples to cheat the adversary's malicious attack classifier. Both AttriGuard \cite{jia2} and MemGuard \cite{jia1} take advantage of the reverse intuition of adversarial examples attack and carefully fabricate fake data instead of directly releasing the original data. It's worth noting that the fake data is not a random choice but an elaborate one. The output vector is modified in a specific way without affecting the data utility or the classification values, then the valid information can be disturbed, achieving the purpose of privacy defense. Nevertheless, this method has its limitations. If a small noise is added to modify the output vector, its performance on the resisting attack is poor; if there is a significant change to the output vector, it will affect data utility. Therefore, seeking proper noises and designing an accurate noise selection strategy is of significance to defending strategies against inference attacks. Besides, to mitigate the information leakage of ML, current work introduces adversarial examples into learning procedures to enhance the robustness of the target classifier against privacy attacks\cite{nasr}. It can be solved into an optimization problem whose goal is to jointly minimize
privacy leakage and maximize prediction accuracy, which is a min-max game that minimizes the classification loss of the objective function and decreases the maximum gain of the membership inference attack. This mechanism also can be brought into solving other defense strategies against privacy leakage.

\subsection{Watermarking Techniques}

Digital watermarking is transferred from multimedia ownership verification to NN  to verify the ownership of the target model, \textit{i.e.}, defending against model extraction attacks. Exploiting the intrinsic generalization and memorization capabilities of neural network architecture allows model defenders to embed and verify crafted watermarks during  training phase and prediction phase. 


It is demonstrated that DNNs have achieved remarkable performance in a good deal of real-world applications \cite{Babenko, wan}. Moreover, numerous excellent deep learning frameworks such as AlexNet \cite{krizhevsky1},  LeNet \cite{lecun}, VGGNet \cite{simonyan1}, ResNet \cite{he} and GoogLeNet \cite{Szegedy}, have helped engineers and researchers to develop systems and research with less effort. Since the target model training task demands a large quantities of labeled data, computing resources, as well as expertise efforts, the trained models can be regarded as an important asset and intellectual property. Therefore, directly sharing trained models without any protection measures induces a new risk in model copyright protection.
In order to protect the copyright and verify the ownership of the trained model, digital watermarking technology was first introduced into deep neural networks training procedure by  Nagai et al. \cite{nagai}. The authors formulated the requirements for an embedding watermark and the corresponding embedding situations. They also introduced three types of attacks against neural networks which these watermarks were placed into should be robust, including fine-tuning, model compression and watermarking overwriting. Furthermore, a concrete framework for embedding watermarking into model parameters was proposed by utilizing a parameter regularizer. The experimental results conducted on two well-known datasets CIFAR-10 and Caltech-101 datasets demonstrated that embedding a watermark via a regularizer without impairing the target model's performance was feasible. Moreover, these embedded watermarks were robust against the three above-mentioned types of attacks.

As we well known, the concept of ``digital watermarking” technique was firstly transferred from multimedia content to the deep neural networks by Nagai et. al \cite{nagai}, aiming at providing the verification mechanism for model ownership. In their design, the watermarks are embedded into the parameters of the target model via the parameter regularizer in the training process. In this way, it is essential for the model owners to access all the parameters of the target model and further extract these watermarks, leading to its white-box constraints. To address the limitations of Nagai et. al \cite{nagai}'s work, 
 Zhang et al. \cite{zhang7} firstly extended the existing DNN watermarking scheme \cite{nagai} into the black-box setting, which supported for model ownership verification with black-box API access. To this end, they proposed three watermarking generation approaches to produce diverse kinds of watermarks and the corresponding framework in order to embed these watermarks to the neural networks. Three types of watermarks including  meaningful content samples, irrelevant data samples and noises are mixed up with the original training data and then fed into the model training process. Then, exploring the model generalization and memorization capabilities, the model learned the patterns of the pre-defined pairs. In the subsequent verification phase, the pre-defined pairs would act as the keys for the model ownership declaration.
The proposed watermarking framework was evaluated on two image recognition benchmark datasets, \emph{i.e.}, MNIST and CIFAR10, revealing that this framework is capable of verifying the ownership of the remote DNN services quickly and accurately without impairing the performance of the original classifiers. Moreover, these embedded watermarks were robust to some malicious attack mechanisms, for instance, model fine-tuning and model pruning. Even though 90\% of model parameters were taken away, all embedded watermarks still had a  high accuracy  (over 99\%). Besides, when model extraction attacks were launched on DNN models embedded with watermarking, there was no watermarks is detected and eliminated.


 \begin{table*}
\begin{center}
  \caption{Comparisons of Defense Strategies Based on Watermarking Techniques}
  \label{tab:commands}
  \begin{tabular}{cccccc}
    \toprule
   Paper, Year & Dataset & Model &  Target information    & Techniques & Defense against attacks \\
\midrule
\multirow{4}{*}{\cite{nagai}, 2018} &  & \multirow{4}{*}{DNN} & \multirow{4}{*}{Model ownership}  & \multirow{4}{*}{ Parameter regularizer} & Fine-tuning,\\
    & CIFAR10,     & & & & Parameter pruning,\\
    & Caltech-101  & &   &  & Distilling,  \\
     &  & &   &  & Model extraction \\
 \midrule

              \multirow{4}{*}{\cite{zhang7}, 2018} &  & \multirow{4}{*}{DNN} & \multirow{4}{*}{Model ownership}  & \multirow{4}{*}{ Training watermarks to DNNs} & Fine-tuning,   \\
    & MNIST, & &   &  & Parameter pruning, \\
      &  CIFAR10 & &   &  & Model inversion attack, \\
          &  & &   &  & Model extraction \\
  \midrule
        \multirow{3}{*}{\cite{jia4}, 2020} & MNIST, & \multirow{3}{*}{CNN, RNN} & \multirow{3}{*}{Model ownership}   & \multirow{3}{*}{EWE}  & \multirow{3}{*}{Model extraction} \\
    & Fashion-MNIST, & &   &  &   \\
      & Google Speech Commands  & &   &   \\
      
  \midrule
              \multirow{3}{*}{ \cite{szyller}, 2019} & MNIST, GTSRB, & Low-capacity DNNs & \multirow{3}{*}{Model ownership}  & \multirow{3}{*}{DAWN}  & \multirow{3}{*}{Model extraction} \\
   & CIFAR10, Caltech,  &  High-capacity DNNs &  &  & \\
    &   ImageNet  & &   &  &   \\
    \bottomrule
  \end{tabular}
  \end{center}
\end{table*}
 
The critical insight behind the existing watermarking scheme \cite{zhang7} is that watermarking uses the untapped model capacity to make the model overfit to some input-output pairs, which are not selected from ordinary task distribution and only known by the model developers. 
However, since watermarks differ in data distribution with the original task distribution, there still exists a flaw in its efficiency. That is to say,  the watermarks can be easily removed through some strategies such as model compression or knowledge transfer. 
Jia et al. \cite{jia4} proposed Entangled Watermarking Embeddings (EWE), which was a feasible method to address this limitation of the AI model watermarking. EWE motivated a target model to extract features that can be utilized to learn how to classify data from the task distribution while predicting the defender's expected output on watermarks. The fundamental intuition was to use the soft nearest neighbor loss \cite{Frosst} to entangle some representations, 
which were extracted from the watermarks and the training data, respectively. In this way, the same subset of parameters was encoded by both watermarks and training data. Therefore, an adversary aiming at extracting the model brought in model's watermarks. In other words, the adversary was forced to learn how to re-generate the defenders' expected output on watermarks. Thus, any attempt to removing watermarks will degrade the model's generalization performance, leading to defeat the original purpose of model extraction attack.
EWE was respectively evaluated on two vision  data sets, \emph{i.e.}, MNIST \cite{lecun1} and Fashion MNIST \cite{xiao}, and an audio dataset, Google Speech Command \cite{PeteWarden}. The  results validated that the defender was capable of claiming the ownership of the target model with $95\%$ confidence after $< 10$ queries submitted by the adversary to steal the model. Besides, the proposed EWE achieved the goal of ownership verification with the moderate cost of model utility loss ($< 1\%$).

Unfortunately, the existing watermarking embedding schemes are not applicable for the black-box access scenario \cite{jia4, nagai, zhang7}, where an adversary trains the surrogate model through collecting the results returned from the API predictions. In this way, the ownership of the surrogate model is not directly verified by using the watermarking techniques, which relies on model owners to embed watermarks into the training procedure. 
To address the above issue, DAWN (Dynamic Adversarial Watermarking of Neural Network) was introduced by Szyller et al. \cite{szyller}, aiming to detect the model extraction attack in a black-box scenario. Rather than being deployed into the training procedure, these watermarks were added to a tiny fraction of queries from the client. Then, the watermarked queries served as a trigger set to decide whether a given model was a surrogate one. That is to say, the model developer can leverage the trigger set to claim the ownership of the targeted surrogate model. However, it inevitably caused two challenges: 1) for the training process, an adversary can choose the training data or alter the training procedure to avoid adopting the watermarks; 2) for the defender, it was a significant challenge to select a perfect trigger set from the whole input query space. In fact, DAWN not only determined whether a given model was a surrogate but also can recognize the malicious client whose queries were used to train the surrogate model. Besides, DAWN randomly selected enough tiny fraction of queries to change the predictions so that the original model's utility for the honest clients would not be degraded. 
In the experiment, it was demonstrated that DAWN was resistant to two existing model extraction attacks:  PRADA attack \cite{juuti} and KnockOff attack \cite{Orekondy}. The satisfactory experimental results showed that the model developers can reliably verify its ownership with the confidence of $> 1-2^{-64}$ with negligible utility loss ($0.03-0.5\%$).

\textbf{Comparisons and Insights.} To build machine learning models for providing Internet users with prediction or decision-making services,  a series of procedures need to be processed. The right training sets should be collected and labeled,  the proper architecture and training parameters are selected to achieve an optimal balance between algorithm accuracy and learning speed, and heavy computational run-time is invested. However, if the intellectual property of this machine learning model is not properly protected, it takes only little effort for a competitor to copy and steal the target machine learning model. An attack will fine-tune it to avoid detection, and then deploy it directly into their products.  Watermarking technique is introduced and designed transferring from the image field to neural networks to defend against this kind of attack, i.e.,  model extraction attack. Table IX presents several current defense strategies by using the watermarking technique in terms of experimental datasets, models, target information, techniques, and types of attacks to resist.

As discussed above, there are two types of embedding watermarking strategies: 1) in the training phase, the defenders exploit the unused model capacity to make the target model overfit to outlier input-output pairs \cite{zhang7}. Then, in the verification phase, when feeding on a specific input to the surrogate model, it outputs the pre-defined prediction. Besides, watermarks also can be regarded as a regularizer and added into the formal objective function. In this way, watermarks are embedded into model parameters without impairing the performance of the classifier \cite{nagai}. Thus, the defender can claim the ownership of the target models;  
2) in the prediction phase,  watermarks are embedded in API predictions without performing any operations on the training procedure. 
It changes the responses of the queries from a malicious attacker who wants to train a surrogate model \cite{szyller}. Also, some related works are devoted to improving the robustness of the watermarking embedding schemes. They entangled the watermarks with the samples from the task distribution to prevent watermarks from being removed  \cite{jia4}. On this occasion, an attacker who intends to remove these embedded watermarks will weaken the performance of the target model with a high probability.


\subsection{Cryptographic Techniques}

Taking active privacy protection measures in machine learning by using cryptographic techniques is an effective way, because they can support meaningful operations in cipher forms to protect private data involved in machine learning algorithms. Thus, designing secure protocols for ML training and prediction phases are urgent requirements for both academia and industries. A generalized cryptographic strategy for privacy defense mechanism is described in
Fig. 4.  It consists of 
two phases: the objective of Phase I is to deliver a large amount of training dataset to the cloud server and train a model with satisfying performance. Multiple users ($i=1,2,...,n$) distribute their private training samples into one or more server(s) to prepare for further training. Then, by leveraging secure multiple computation or homomorphic encryption techniques, a well-generalized machine learning model $\emph{W}$ is intensively trained and released without exposing the truth values of the training dataset. In Phase II, there exists a user who desires to query the target model for obtaining classification results $y$ without revealing the content of query data $b$. Additionally, model parameters are also deemed as intellectual property due to its confidential training data and intensive computational resources to train the classifier. Hence, to protect the private information of both model parameters and query samples, customized cryptographic tools are exploited to design secure protocols for model prediction services. So far, we have provided the general framework of privacy-preserving machine learning by utilizing cryptographic techniques. In the following, we will review some representative works in the specific scenarios.

\begin{figure*}[!thbp]
	\centering
	\includegraphics[width=.7\textwidth]{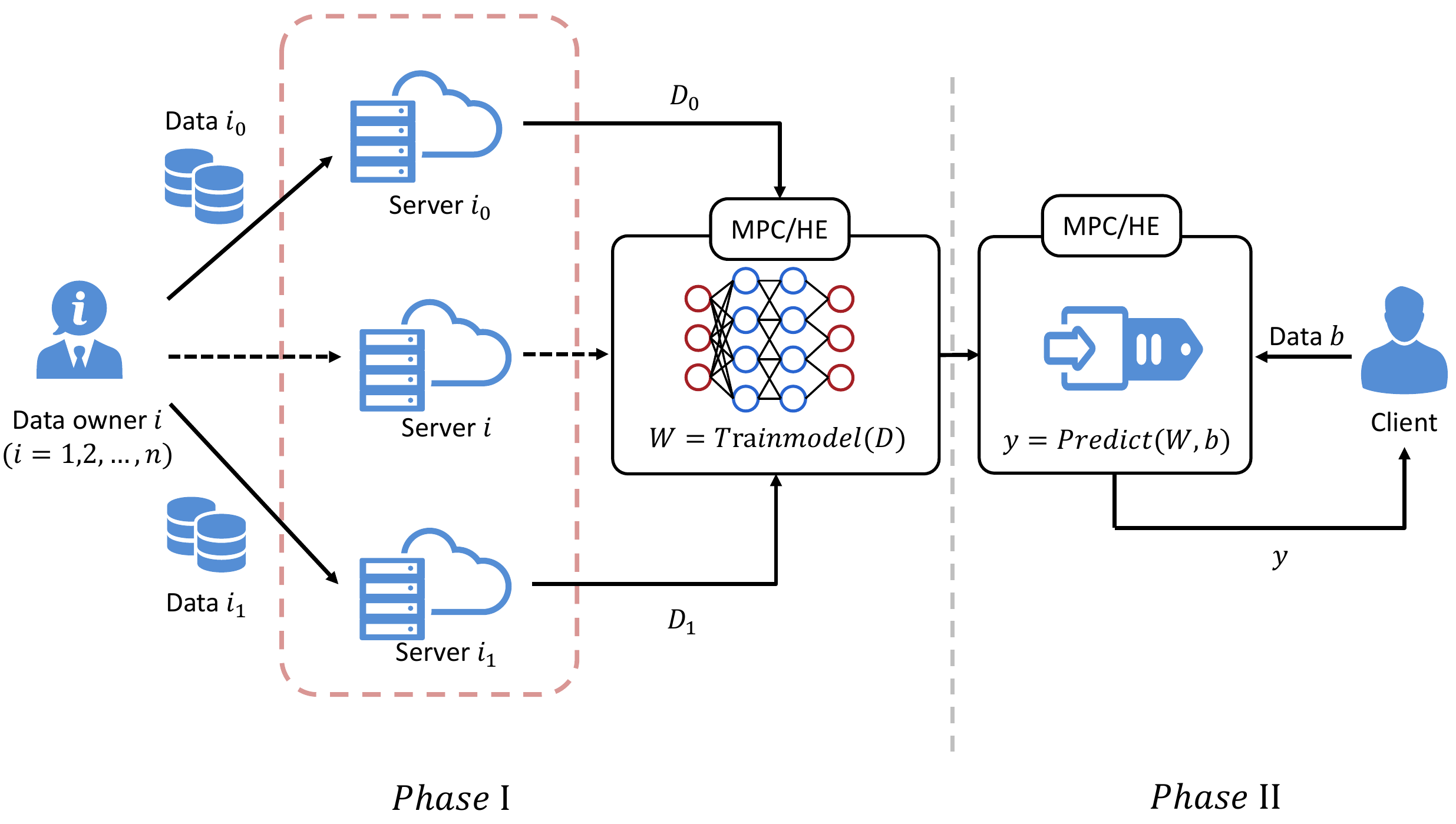}
	\caption{A generalized cryptographic strategy for privacy defence}
	\label{figure_system_model}
\end{figure*}

The wide use of mobile devices generates rich but sensitive data in real-world. Large-scale sensitive data collection leads to more accurate machine learning models yet poses sensitive data at risks. It is mainly owing to the training data collection without any blinding or encryption processing. Moreover, the federated machine learning where multiple users locally train and jointly build a shared model significantly improves the model performance. However, the existing adversarial attacks such as model inversion attacks can still steal victim user's training data through observing gradient updates, which can be viewed as confidential information \cite{hitaj}. Hence, secure aggregation protocols are urgently required to design in order to address the privacy issues of information leakage in both data manipulation or model training phases.
Bonawitz et al. \cite{bonawitz} focused on mobile devices where communication was costly, and the workers' dropout occurred frequently. They proposed two secure aggregation variant protocols which were low-communication overhead and robust to failures. The first protocol was more effective  and secure against semi-honest attackers while the latter one can guarantee privacy against active adversaries with some extra cost. The main idea behind their design was double masking. Based on the threshold secret sharing, two blinding factors were leveraged to conceal the original input data and distributed the remaining workers. The first blinding factors were generated via the Diffie-Hellman protocol and can be canceled each other out in the aggregation phase. The second blinding factors were chosen randomly by each worker and required to be reconstructed by the remaining workers in the aggregation phase when workers dropout. Performance evaluations demonstrated its efficiency, i.e., the communication cost and the running time remained low level on the large datasets. In terms of  $16$-bit input values, the protocol offered $1.73\times$ communication expansion for $2^{10}$ workers and $2^{20}$-dimensional vectors, and $1.98\times$ expansion for $2^{14}$ workers and $2^{24}$-dimensional vectors compared to the input data in plaintexts.


Mohassel et al. \cite{mohassel} focused on a two-server scenario where data contributors distributed samples to two non-colluding service providers. In the setup stage, the data owners delivered their data to two service vendors by leveraging encryption or secret-sharing methods. In the following training stage, two cloud servers collaboratively trained classifiers on united data without obtaining any extra knowledge beyond the 
obtained target classifiers. 
The architecture of the NN consisted of linear functions and an activation function. That is to say, all calculation operations involved in both forward and backward propagation included additions, multiplications, activation function and its corresponding partial derivative. The computation of linear functions consisted of online and offline phases. More concretely, the first phase focused on updating the coefficients by using the shared values while the second phase was primarily comprised of triple multiplication generation. To further improve its efficiency, the mini-batch and vectorization techniques were utilized for training. For the activation function, the RELU function was used to replace the softmax function in order to compute efficiently. To evaluate RELU  and its derivative,  Yao sharing \cite{yao} was used. Besides, the RELU function and its derivatives enabled to be computed together in one round, and then the  latter was utilized in the backward propagation.
As the experimental results showed, the proposed protocols dramatically minimized the difference between privacy-preserving training and training in plaintext forms. For instance, the total run time of the privacy-preserving procedure for learning a NN with $266$ neurons and $3$ layers was $21,000$s. Besides, the proposed protocol can be divided into the online and offline procedures, which was outperforms compared to the training in plaintext without the offline phase. For instance, in the LAN setting, the neural network training took $653$s. Moreover, the proposed activation function can achieve a competitive accuracy at $93.4\%$ compared to that of the original softmax function at $94.5\%$ while implementing privacy-preserving NN  training.

Different from two-server scenario \cite{mohassel}, the setting considered in \cite{mohassel1} is a multi-server model where users sent their sensitive data in ciphertext form to several service providers who conducted the learning protocol on the united data sets. 
Although new MPC techniques and implementations have dramatically boosted their performance, it is still constrained to the two-server model while no  benefiting from these speedups. The first reason is that the-state-of-art MPC \cite{Nikolaenko1, nikolaenko} methods are only appropriate for computations over a ring while the samples involved in machine learning are decimal values. The second reason is that neural networks require 
transferring back and forth between arithmetic and non-arithmetic computation operations freely. 
However, converting different sharing forms instantiated by standard ways is costly and inefficient.
Mohassel \cite{mohassel1} designed a new mixed framework for neural networks, which consisted of five building blocks. In details, a new fixed-point multiplication scheme for any number of parties
was designed, resulting in considerable improvement in throughput, latency and communication. Matrix multiplication is the primary computation for neural network. Then, vectorized multiplication among all parties was designed to implement a linear combination of multiplication terms. Besides, share conversions between all types of sharing were designed and optimized to facilitate efficient conversions. Performing on mixed representations can be more efficient, then the authors designed a mixed-protocol to perform multiplication operations for different representations. Finally, they gave an efficient computation protocol on polynomial piece-wise functions.
The framework was implemented for training and inference on NN models in the semi-honest setting. Compared to the two-party solution of SecureML \cite{mohassel}, the proposed solutions were up to 5500 times faster with the same accuracy $94\%$ when training a neural network model. Besides, the model can produce a handwriting prediction in $10ms$ while the advanced Chameleon \cite{riazi} scheme taking $2700ms$.

However, most works for training \cite{hesamifard, mohassel, mohassel1, wagh} and prediction \cite{bourse, gilad-bachrach, sanyal} on encrypted data mainly focus on optimizations for either the internal machine learning algorithms or the applied cryptographic techniques discretely. Hence, designing an optimized ML model alongside a customized secure computation protocol will facilitate the performance of privacy-preserving ML.
Agrawal et al. \cite{agrawal} presented a new protocol to optimize DNNs as well as  a specific crafted  secure two-party protocol, named QUOTIENT. Similar to several previous work \cite{gascon, kilbertus, mohassel, nikolaenko}, the training process was outsourced to two non-colluding servers. The authors mainly made the following modifications to achieve the designed objective. Firstly, the weight parameters $\textbf{W}$ were internalized during the forward and backward propagation, \emph{i.e.}, $\textbf{W}\in \left\{-1, 0, 1\right\}$, allowing to multiply $\textbf{W}$ to be expressed as oblivious transfers. Secondly, an MPC-friendly quantization function was built for the updates, substituting a biased coin flip and truncation operations with a saturation-free quantization, without 
any loss of accuracy. Finally, in the backward pass normalization division operation, they replaced the \emph{closest-power-of-two} with \emph{next-power-of-two}. This minor modification made a great improvement in further circuit implementation leveraged in security protocols.
The proposed QUOTIENT was performed on six different real datasets.
In terms of the secure training phase, compared with the advanced work, QUOTIENT achieved $99.38\%$ classification accuracy on MNIST, improving $6\%$ over SecureML's \cite{mohassel} error rates upon convergence. For prediction phase, compared to \cite{mohassel} the proposed protocol was $13\times$ faster for a single prediction and $7\times$ faster for batched predictions over LAN while achieving $3\times$ speed-up for a single prediction and $50\times$ for batched predictions over WAN.
Hence, QUOTIENT makes 2PC (secure two-party computation) neural network training practical.

Recently, Liu et al. \cite{liu1} concentrated on exploring the privacy issue in prediction phase, \emph{i.e.,} after each prediction, the users acquired  nothing about the trained classifier except for the final prediction values while the server learned nothing about the users' query data. 
They presented an oblivious protocol for transforming an arbitrary NN into an oblivious one  without considering how models were trained, named MiniONN. The neural networks commonly consist of routine operations including \emph{linear transformations, popular activation functions} and \emph{pooling operations}. Thus, Liu et al. \cite{liu1} designed several oblivious protocols for each building block.
Especially, before conducting a linear transformation operation, a pre-computation phase was adopted to generate dot-product triplets. It was typically generated by leveraging homomorphic encryption or oblivious transfer. Next, oblivious linear transformations, oblivious activation functions and oblivious pooling operations protocols were designed, respectively. In addition, based on the security of dot-product triplet generation, all operations mentioned above can be performed by utilizing a secure two-party protocol. Hence, the security requirements of them were satisfied.
MiniONN was conducted on three real datasets, \emph{i.e.,} Handwriting recognition (MNIST), image classification (CIFAR-10) and language modeling (PTB). Compared to Secure ML \cite{mohassel}, MiniONN gained better performance in latency and accuracy. Compared to CryptoNets \cite{gilad-bachrach}, MiniONN achieved a 230-fold reduction in latency and 8-fold reduction in message sizes without any accuracy loss. Besides, the experimental results showed that the performance of the ONNs in MNIST and PTB was reasonable while the ONNs in CIFAR-10 is underperformed.


Secure outsourced matrix computation based on HE makes a significant contribution to privacy-preserving machine learning. In general, there are three outsourced model prediction scenarios depending on who provides the model and the  query data. 
The first one is that the data owner learns a target classifier and makes it open access through a platform to offer prediction function by inputting encrypted query data; the second one is that model producer encrypts the classifier and submits it to the service platform for prediction services; the third one is that the server provider trains a classifier on the large encrypted datasets and uses it to predict on the  encrypted samples.
Several works only focus on applying a plaintext classifier to encrypted data (the first scenario)  \cite{bourse, chabanne, gilad-bachrach},  while the second and the third settings have not been explored yet.
Jiang et al. \cite{jiang} proposed several efficient matrix operations by combining HE-based operations on packed ciphertexts including Single Instruction Multiple Data  arithmetic, scalar multiplication and slot rotation. This solution was under the condition that a ciphertext can encrypt $d^{2}$ plaintext slots, but it also can be used to support matrix computation of any arbitrary size. When applying secure matrix multiplication into CNNs training, the data owners encrypted their image data while the model provider encrypting multiple convolution kernels' values of convolution layers and weights of FC layers, and then sending them to the cloud server. Hence, the problem of CNNs prediction on ciphertexts  was transferred into the problem of matrix computation with a HE system.
As for secure outsourced matrix computation, its implementation took $9.21s$ to multiply two encrypted matrices of order $64$ and $2.56s$  to transpose a square matrix of order $64$. When applying it to the convolutional neural network (CNN) on the MNIST dataset, the experimental results demonstrated that it takes $28.59s$ to compute ten likelihoods of 64 input images at the same time, amounting to an amortized rate of $0.45s$  each image.


Secure computation might be used to alleviate the privacy issues in deep learning, for instance, MPC and HE \cite{gentry}.
Although significant progress has been made in MPC techniques, the expensive communication overhead is still the large obstacle of MPC implementation. Also, HE refers to a cryptosystem that allows someone to perform meaningful operations on encrypted data without conducting decryption operations. However, the most existing HE schemes only support operations on ciphertexts encrypted with one secret key. Hence, supporting secure computation among several data contributors requires a novel HE scheme with multiple different secret keys. Then, 
Chen et al. \cite{chen} proposed an RLWE-based cryptosystem to achieve homomorphic multi
plication of multi-key variants BFV \cite{brakerski, fan} or CKKS \cite{cheon} schemes, consisting of tensor product and relinearization steps. Especially, the authors designed two relinearization methods, depending on the previous GSW ciphertext extension \cite{mukherjee} and directly linearizing each entry of an extended ciphertext. Besides, in order to avoid expensive high-precision arithmetic in implementation, they used an RNS-friendly decomposition method \cite{bajard, cheon} for relinearization. Hence, the proposed MKHE can be explored to secure online classification services scenario where the new samples and the target model were encrypted with two distinct secret keys.
The performance evaluations showed that implementing the MKHE scheme on the MNIST dataset took $1.8s$  to classify an encrypted image, which was fed into one convolutional layer followed by two fully connected layers. It indicated that the MKHE scheme was useful and practical when applied into oblivious neural network inference.

\begin{table}
\setlength{\tabcolsep}{1mm}{
  \caption{Comparisons of Defense Strategies Based on Cryptographic Techniques}
  \label{tab:commands}
  \begin{tabular}{cccc}
    \toprule
   Paper, Year & Dataset & Model & Techniques  \\
\midrule
 \cite{bonawitz}, 2017 & - & Generalized & SS; Blinding   \\

\midrule
        
        \cite{mohassel}, 2017 & MNIST, Gisette & NN & 3PC \\

\midrule
\cite{mohassel1}, 2018 & Synthetic datasets,  MNIST & NN, CNN & SS; 3PC   \\
      \midrule

           \multirow{3}{*}  {\cite{agrawal}, 2019} & MNIST, MotionSense,  & \multirow{3}{*}{DNN} &   \multirow{3}{*}{2PC} \\
           &  Thyroid,  Breast cancer, Skin  &  &    \\
& Cancer MNIST, German credit, &  &    \\

        \midrule
              \cite{liu1}, 2017 & MNIST,  CIFAR-10,  PTB & Generalized & OT; 2PC \\

  \midrule
          \cite{jiang}, 2018 & MINST  & CNN   & HE\\

        \midrule
           \cite{chen}, 2019  &MNIST & CNN  & MKHE \\

    \bottomrule
  \end{tabular}}
\end{table}

\textbf{Comparisons and Insights.} Table X summarizes all defense strategies based on cryptographic techniques reviewed in this survey. It is not difficult to find that secure multiparty computation and homomorphic encryption are two promising cryptographic tools for designing privacy-preserving ML schemes. During the model training procedure, secure multiparty computation allows the data owner and the model developer to securely interact with each other, aiming at learning a well-trained model. When deploying ML models on the cloud-based service platform, a secure two-party computation mechanism is used to guarantee the model privacy of the model producer and the query data privacy of the model consumer simultaneously. Especially,  in the collaborative learning scenario, two or more cloud servers  \cite{mohassel, mohassel1} are introduced to cooperatively train a common classifier. Multiple data providers distribute their data to several non-colluding cloud servers and then secure multiparty computation is leveraged to learn a standard classifier among these servers. That is to say, a technological breakthrough in the basic secure multiparty computation algorithm accelerates the progress of privacy-preserving machine learning.

Homomorphic encryption enables the participant to perform arithmetic operations on ciphertexts but retains its meaningful computations. Thus, homomorphic encryption can be deployed to protect each participant's data privacy during ML model training and prediction procedures. Since ML training and prediction procedures involve several interactions and iterations among multiple entities, efficient multi-key homomorphic encryption schemes are designed to deal with privacy leakage issues \cite{chen}. 
Although using HE solutions has many advantages on ciphertext computation, there still exist some limitations. The first one is that almost all HE schemes use integers in the information space, so the input data items need to be converted to integers before encrypting. The second limitation is the size of the ciphertext.
The size of the message is greatly increased with the continuous encryption operations.  After each encryption operation, the amount of noise in the ciphertext increases, and more noises will lead to an incorrect decrypted result. Therefore, we should always keep the noise levels below the pre-defined threshold.

\section{Challenges and Future Work}
In the previous section, we have presented recent research achievements regarding privacy attack and protection in cloud-based neural networks. Nevertheless, several types of attacks rely on some strong assumptions, so that these attacks are infeasible in real-world scenarios. Besides, the secure computation used for privacy protection is often based on computing-intensive cryptographic tools or low-precision confusion technique such as differential privacy, seeking the trade-off among computation overhead, communication overhead and model precision. Hence, in this section, we will discuss the challenges in front of researchers who are interested in this filed. After that, we propose some future research directions related to privacy attacks and its corresponding countermeasures alongside the pipeline of neural networks.

\subsection{Privacy attack}

\emph{Membership leakage of explaining machine learning models.} Although the neural network has shown its potentials in a variety of classification tasks,  the lack of its transparent results in its limited application in security-critical domains \cite{junyang2020}.
As a result, plenty of current works are attempted to provide interpretable ML models to explain the classification results \cite{guo, Lundberg, Ribeiro}. Recently, it has been demonstrated that explanations of machine learning can leak the membership status of individual training data point \cite{shokri3}. This finding reveals that releasing the transparent reports of machine learning will lead to privacy leakage of the training dataset.
However, there are still some interesting open issues to be addressed. Firstly, the target model investigated is restricted to simple ML model such as logistic regression, and extending the similar attacks to other more complicated models remains an unsolved issue. Secondly, both safe and useful explanations are required in many safe-critical applications. Therefore, designing safe transparency reports is another effective way to deal with privacy risks.

\emph{Stealing model parameters without knowing the ML algorithm.} It has been demonstrated that stealing hyperparameters via black-box attack under the assumption that the ML algorithm is known is feasible \cite{wang}. Unfortunately, the proposed current attack is unsuitable for the scenario where the adversary aims at collaboratively stealing the Machine learning algorithm and the hyperparameters. In reality, jointly plagiarizing both the ML algorithm and the hyperparameters may be infeasible or even impossible. More precisely, a neural network with a hyperparameter \emph{A} generates a classifier $M_{A}$. For the identical training data set, logistic regression with a hyperparameter \emph{B} generates a classifier $M_{B}$. If the model parameter of $M_{A}$ is identical to that of $M_{B}$, we can not distinguish between the neural network with hyperparameter \emph{A} and the logistic regression with hyperparameter \emph{B}. Hence, it is well worth exploring to steal ML algorithm and hyperparameters in future work simultaneously.


\emph{Privacy risk of securing machine learning.} Investigating the connection of security domain and privacy domain indicates that the robust training approaches can result in membership inference privacy leakage. It has been revealed that the privacy leakage is related to the target classifier's robustness generalization, its adversarial perturbation constraint, and its capacity. It is worth to note that the failure of privacy leakage of robustness generalization may be thanks to adding wrong or inappropriate distance constraints such as $l_{p}$, which are utilized to adversarial model training. While $l_{p}$ perturbation constraints have been extensively applied in adversarial machine learning \cite{Biggio, Goodfellow, Jacobsen, Madry, wong}, it still exists some limitations. Moreover, it has been demonstrated that $l_{p}$ distance has no inevitable connection with image semantics or visual perception \cite{Sharif}. In addition to membership inference attack, the robust model obtained via adversarial machine learning also suffers from another type of adversarial example, which changes the semantics of the image but keeps the predictive model results unchanged \cite{Jorn-Henrik}. Thus, in future work, the question of whether the privacy-robustness conflict is fundamental remains as an open problem for the research community. Furthermore, meaningful perturbation constraints to seek the 
straightforward tension between model robustness and model privacy continue to be an essential question.

\emph{Designing stronger and more applicable membership inference attack.} 
MInf was firstly presented by Shokri et al. \cite{shokri1}, which relies on many strong assumptions, such as leveraging several so-called shadow models, prior information of the target classifier's architecture and the same data distribution of the target classifier's training samples. Then, all these strong assumptions were gradually relaxed, and three corresponding types of membership inference attacks were proposed in \cite{salem}. In particular, the strongest adversary relaxes all the assumptions of training any shadow model, prior knowledge of the target model and data distribution. Moreover, it was implemented as unsupervised training without knowing the class label of the target data. However, while there are many predominant advantages over other attacks, this type of adversary does not work well in some data domains, where the bounds for input features are not explicit. Hence, it is necessary to remove the limitations of the state-of-the-art work  \cite{salem} and expand the membership inference attack's application for all kinds of data sampels. In future work, designing stronger and more applicable membership inference attack against more complex DNN models will be promising and useful.

\subsection{Privacy defense}

\emph{Efficient and Mixed Secure Computation Protocol for Machine Learning Applications.}
Cloud-based secure computation can be used to improve the efficiency and accuracy of machine learning. Considering a scenario where one or more untrusted cloud servers are employed to reduce local computational resources, privacy risk of individually outsourced data notably become a severe problem to be dealt with. A range of cryptographic tools, such as secure multi-party computation, secret sharing, Yao’s garbled circuit protocol, fully-homomorphic encryption, Goldreich-Micali-Wigderson protocol, are exploited in ML-based outsourcing computation to tackle with privacy leakage issues. Significant research effort demonstrates that the single technique leads to low performance, saving on communication costs at the sacrifice of expensive computation operations, or vice versa. To address this challenge, a promising research direction in future work is to skillfully combine multiple techniques, designing mixed-protocol secure computation protocols for machine learning algorithms, which can achieve a better trade-off among computation, communication while ensuring the security of designed mixed protocols.

\emph{Collaborative Learning against Malicious Participants.} 
The basic idea behind federated learning is to collaboratively train more accurate models from multiple datasets while preserving each distributor's  
original training set. In particular, there exists a parameter server provider who is responsible for assigning learning tasks, aggregating the local trained model's parameters and updating a global target model. Plenty of current works focus on federated learning scenarios where all the providers are honest to follow the protocols. However, real-world applications often encounter such situations that some of the nodes might be unreliable or malicious, which means that some data providers might not behave as intended. Traditional federated learning becomes fragile when attacked by such Byzantine adversaries. Moreover, it has been proven that the parameter updates leak unintended information about participants' training data. Adversaries are motivated to develop passive and active inference attacks to exploit this leakage. Therefore, how to construct a federated learning system that is secure against Byzantine adversaries while preventing parameter updates leakage becomes an interesting problem.

\emph{Designing principles of employing ML algorithms for mitigating model inversion attack.}
DNNs have an excellent performance in a great deal of applications, making them successful in becoming a commodity. Several ML frameworks and services are available to data owners who are not skilled in ML training but desire to obtain predictive models from their data. Nevertheless, it is shown that malicious machine learning algorithms provided by vendors can produce models that have good performance in prediction accuracy as well as generalizability. This attack can cause causal to leak a great deal of information related to their training data samples only under the assumption of black-box access \cite{song}. Hence, in this way, ML algorithms from outside can not be directly deployed into model training, especially for sensitive data domains.
Whenever they adopt someone else's ML training algorithms or provide the trained model as a service, it is necessary to see the source code and understand its inner working principle. In future work, we should devote ourselves to designing the principle for ML training frameworks. This principle can guarantee that the trained model only captures as much information about its training data samples as it requires for training task and nothing more. Then, how to formalize this principle, how to implement these training methods into practice, and how to evaluate the safety, the effect and the practicality of these methods are left as open questions for the research community.


\emph{Seeking the trade-off among privacy, utility and efficiency.}
It is well-known that defense measures are taken in outsourcing computing, such as blinding, data perturbation, and data encryption, perform well on protecting raw datasets. Whereas, it inevitably sacrifices other critical indicators \textit{e.g.}, utility and efficiency. Moreover, the dominating influences on its efficiency caused by privacy defense mechanisms constitute of computation cost and communication overhead. 
While cryptographic techniques including FHE and SMC  are capable of performing meaningful operations in ciphertexts, it brings expensive and computation-intensive ML mode. Also, compared to conventional centralized training paradigm, secure outsourcing computing involved in neural network training induces substantial interactive rounds, leading to a considerable amount of communication overhead. In terms of defense mechanism based on differential privacy, although it incurs relatively straightforward calculations, there exists a distinct trade-off between privacy and utility, \textit{i.e.}, the stronger privacy guaranteed, the less model utility it will be. Thus, a promising research direction is to seek the nash equilibrium point under three critical indicators comprising of privacy, utility, and efficiency. Furthermore, establishing privacy evaluation mechanism facilitates better awareness regarding the three indicators above and further improves the performance of outsourcing computing applied in a machine learning paradigm.

\emph{Designing customized privacy protection for end-user.}
A large volume of datasets generated by various Internet users or mobile devices contributes to a more accurate model, resulting in the rise of collaborative learning. In these settings, what they have in common is that there exist a plurality of data contributors or data providers. Different from traditional federated learning scenarios where all participants' datasets are preserved, we focus in particular on some scenarios that only a few fractions of participants or some specific attributes need to be protected. As a consequence, it is necessary to design a personalized privacy protection mechanism relying on divergent privacy requirements of different participants. In other words, conventional privacy protection measures such as differential privacy, homomorphic encryption and secure multi-party computation techniques can not be directly deployed into ML paradigm any longer. Therefore, significant efforts need to be made to explore a dynamic perception of privacy requirements and the corresponding allocation of customized privacy budget in designing secure computation protocols.
In this way, the target model obtained from collaborative learning can achieve high prediction accuracy as well as satisfy different demands for privacy protection required by various participants.

\section{Conclusion}

This survey presents the most recent findings of privacy leakage and protection regarding the cloud-based neural network. A general methodology is extracted and sketched from collecting and reviewing the related literature in the past five years.
We systematically and comprehensively divide the pipeline of neural network learning and application into three phases, i.e., data manipulation, training, and prediction. Furthermore, privacy attack and defense methods are thoroughly reviewed following the thread of neural networks' workflow. More importantly, we introduce the challenges and future work, which will be helpful for researchers to continue promoting the competitions between privacy attackers and defenders in the cloud-based neural network.

\appendices

\ifCLASSOPTIONcaptionsoff
  \newpage
\fi

\end{document}